\documentclass[twocolumn,amsmath,showkeys,amssymb,superscriptaddress,nofootinbib,longbibliography,twocolumn,floatfix,aps,prx,10pt]{revtex4-2}
\usepackage[utf8]{inputenc}
\usepackage[english]{babel}
\usepackage{braket}
\usepackage{bm}
\usepackage{natbib}
\usepackage{amsmath, amssymb}
\usepackage{graphicx}
\usepackage{dcolumn}
\usepackage{xcolor}
\usepackage{empheq}
\usepackage{xr}
\usepackage[colorlinks=true, allcolors=blue]{hyperref}
\def\be{\begin{equation}}
\def\ee{\end{equation}}
\def\bc{\begin{center}}
\def\ec{\end{center}}
\def\bea{\begin{eqnarray}}
\def\eea{\end{eqnarray}}

\newcommand{\gin}[1]{\textcolor{cyan}{\textbf{#1}}}

\newcommand{\Avg}[1]{\left\langle{#1}\right\rangle}

\newcommand\mymapsto{\mathrel{\ooalign{$\rightarrow$\cr%
  \kern-.15ex\raise.275ex\hbox{\scalebox{1}[0.522]{$\mid$}}\cr}}}

\newcommand{\real}{\mathbb{R}}

\newcommand{\vX}{\mbox{\boldmath $X$}}
\newcommand{\mL}{\mathbf{L}}

\newcommand{\mI}{\mathbf{I}}
\newcommand{\di}{\displaystyle}

\usepackage{pdfpages} 
\usepackage{pgffor} 

\makeatletter
\AtBeginDocument{\let\LS@rot\@undefined}
\makeatother

\begin{document}

\title{Mining higher-order triadic interactions}

\author{Marta Niedostatek$^*$}
\affiliation{School of Mathematical Sciences, Queen Mary University of London, London, E1 4NS, United Kingdom}
\affiliation{The Alan Turing Institute, The British Library, London, NW1 2DB, United Kingdom}

\author{Anthony Baptista$^*$}
\affiliation{School of Mathematical Sciences, Queen Mary University of London, London, E1 4NS, United Kingdom}
\affiliation{The Alan Turing Institute, The British Library, London, NW1 2DB, United Kingdom}
\affiliation{Cancer Bioinformatics, School of Cancer and Pharmaceutical Sciences, Faculty of Life Sciences and Medicine, King’s College London, London, WC2R 2LS, UK}

\author{Jun Yamamoto}
\affiliation{Department of Network and Data Science, Central European University, Vienna 1100, Austria}

\author{J\"urgen Kurths}
\affiliation{Potsdam Institute for Climate Impact Research,
14473 Potsdam, Germany}
\affiliation{Institute of Physics, Humboldt University of Berlin, 12489 Berlin, Germany}

\author{Ruben Sanchez Garcia}
\affiliation{The Alan Turing Institute, The British Library, London, NW1 2DB, United Kingdom}
\affiliation{School of Mathematical Sciences, University of Southampton, Southampton SO17 1BJ, United Kingdom}

\author{Ben MacArthur}
\affiliation{The Alan Turing Institute, The British Library, London, NW1 2DB, United Kingdom}
\affiliation{School of Mathematical Sciences, University of Southampton, Southampton SO17 1BJ, United Kingdom}
\affiliation{Faculty of Medicine, University of Southampton, Southampton SO17 1BJ, United Kingdom}

\author{Ginestra Bianconi}
\affiliation{School of Mathematical Sciences, Queen Mary University of London, London, E1 4NS, United Kingdom}
\affiliation{The Alan Turing Institute, The British Library, London, NW1 2DB, United Kingdom}

\begin{abstract}
Complex systems often involve higher-order interactions which require us to go beyond their description in terms of pairwise networks. Triadic interactions are a fundamental type of higher-order interaction that occurs when one node regulates the interaction between two other nodes. Triadic interactions are found in a large variety of biological systems, from neuron-glia interactions to gene-regulation and ecosystems. However, triadic interactions have so far been mostly neglected. In this article, we propose  {the Triadic Perceptron Model (TPM)} that demonstrates that triadic interactions can modulate the mutual information between the dynamical state of two linked nodes. Leveraging this result, we formulate the Triadic Interaction Mining (TRIM) algorithm to extract triadic interactions from node metadata, and we apply this framework to gene expression data, finding new candidates for triadic interactions relevant for Acute Myeloid Leukemia.
Our work reveals important aspects of higher-order triadic interactions that are often ignored, yet can transform our understanding of complex systems and be applied to a large variety of systems ranging from biology to climate.
\end{abstract}

\maketitle

\def\thefootnote{*}\footnotetext{These authors contributed equally to this work}\def\thefootnote{\arabic{footnote}}

\section{Introduction}

Higher-order networks \cite{Bianconi2021,Battiston2021,battiston2020networks,torres2021and,Bick2022} are key to capturing many-body interactions present in complex systems. Inferring higher-order interactions~\cite{Young2021,Contisciani2022,malizia2024reconstructing,musciotto2021detecting,delabays2024hypergraph,lizotte2023hypergraph} from real, pairwise network datasets is recognised as one of the central challenges in the study of higher-order networks \cite{Battiston2021,rosas2022disentangling}, with wide applicability  across different scientific domains, from biology and brain research~\cite{rosas2019quantifying,stramaglia2021quantifying,olbrich2015information} to finance \cite{massara2016network,tumminello2005tool}. 
Mining higher-order interactions from the exclusive knowledge of the pairwise networks typically involves generative models and Bayesian approaches based on network structural properties \cite{Young2021,wegner2024nonparametric,Contisciani2022,musciotto2021detecting,lizotte2023hypergraph}. Note, however, that when the inference is performed on the basis of the knowledge of the nodes' dynamical states \cite{delabays2024hypergraph,malizia2024reconstructing}, inferring higher-order interactions also requires dynamical considerations.

 Triadic interactions \cite{sun2023dynamic} are a fundamental type of signed higher-order interaction that are gaining increasing attention from the statistical mechanics community ~\cite{sun2023dynamic,millan2023triadic,sun2024higher,herron2023robust,kozachkov2023neuron,nicoletti2023information}, since they are  not reducible to hyperedges or simplices. A triadic interaction  occurs when one or more nodes regulate the interaction between two other nodes. The regulator nodes may either enhance or inhibit the interaction between the other two nodes. Triadic interactions are known to be important in various systems, including: ecosystems \cite{Bairey2016,Grilli2017,Letten2019} where one species can regulate the interaction between two other species; neuronal networks \cite{Cho2016}, where glial cells regulate synaptic transmission between neurons thereby controlling brain information processing; and gene regulatory networks \cite{Wang2009, Giorgi2014}, where a modulator can promote or inhibit the interaction between a transcription factor and its target gene. 
There is mounting evidence that triadic  interactions can induce collective phenomena and/or modulate dynamical states that reveal important aspects of complex system behavior~\cite{sun2023dynamic,millan2023triadic,sun2024higher,herron2023robust,kozachkov2023neuron,nicoletti2023information,gao2025triadic,iskrzynski2025pangraphs}. An important advance  in this line of research is triadic percolation~\cite{sun2023dynamic,millan2023triadic,sun2024higher}, a theoretical framework that captures the non-trivial dynamics of the giant component. Moreover, recent results demonstrate that  triadic interactions can have significant effects on  stochastic dynamics \cite{nicoletti2023information} and learning ~\cite{herron2023robust,kozachkov2023neuron}. 
 However, despite the increasing attention that higher-order interactions are receiving, the detection of triadic interactions from network data and node time series, is an important scientific challenge  that has not been thoroughly explored~\cite{Kenett2015,Zhao2016,Wang2009}.

In this article,  we  formulate the {Triadic Perceptron Model (TPM) in which continuous node variables are affected by triadic interactions. Based on the insights gained by investigating this model, we  propose an information theoretic approach, leading to the Triadic Interaction Mining (TRIM) algorithm, for mining triadic interactions.}
The TPM provides evidence of the  mechanisms by which a triadic interaction can induce a significant  variability of the mutual information between two nodes at the end-points of an edge.
The TRIM algorithm leverages this finding and mines triadic interactions using knowledge of the network structure and the dynamical variables associated with the nodes. The significance of each putative triadic interaction is then validated by comparison with two distinct null models. 

In this way, the TRIM algorithm can go beyond monotonicity assumptions regarding the functional form of the regulation of the two linked nodes by the third node (which is at the foundation of previously proposed  methods \cite{Wang2009}) allowing  for broader applications.  Significant node triples are also associated with an normalized entropic score function $S\in [0,1]$ that quantifies the spread of the conditional joint distribution functions of the variables at the ends of the regulated edge.
We test the TRIM algorithm  on the benchmark TPM, demonstrating its efficiency in detecting true triadic interactions. We also use the TRIM model to mine triadic interactions from gene-expression, to identify `trigenic' processes \cite{Kuzmin2018}. We demonstrate that the TRIM algorithm is  able to detect known interactions as well as propose a set of new candidate interactions that can then be validated experimentally.

\begin{figure*}[!ht]
    \centering
    \includegraphics[width=16cm,keepaspectratio]{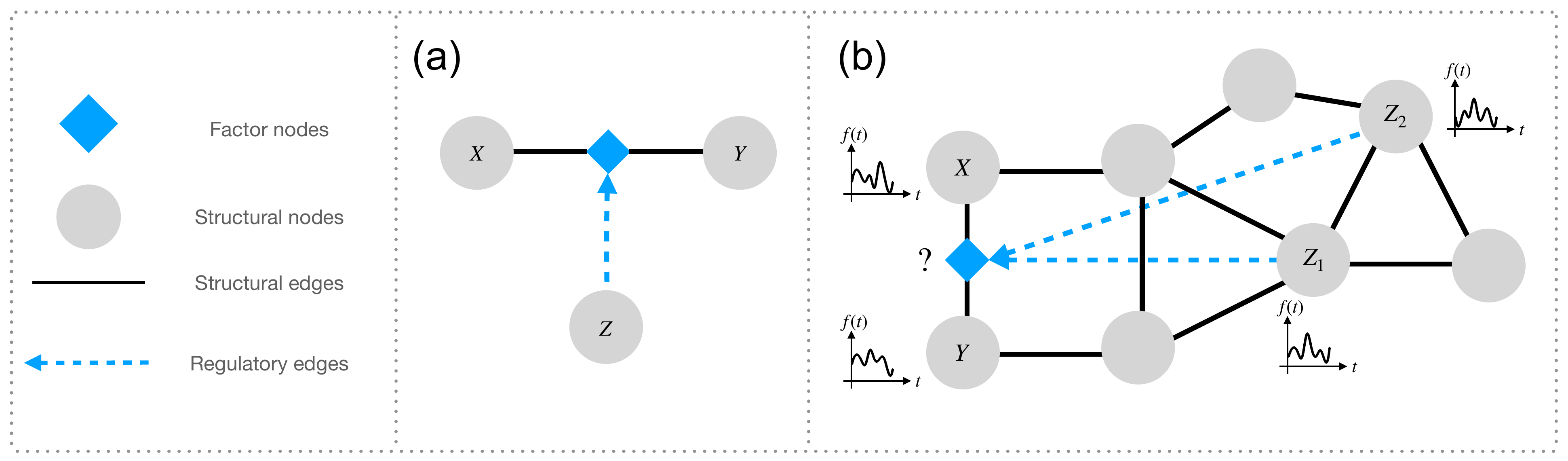}
    \caption{(Panel a) A triadic interaction occurs when a node $Z$, called a \emph{regulator node},  regulates (either positively or negatively) the interaction between two other nodes  $X$ and $Y$.  The regulated edge can be conceptualized as a \emph{factor node} (shown here as a cyan diamond). (Panel b) A network with triadic interactions can be seen as a network of networks formed by a simple structural network and by a bipartite regulatory network between regulator nodes and regulated edges (factor nodes).}
\end{figure*}

\section{Triadic interactions}
A triadic interaction occurs when one or more nodes modulate (or regulate) the interaction between two other nodes, either positively or negatively.
A \emph{triadic interaction network} is a heterogeneous network composed of a structural network and a regulatory network encoding triadic interactions (Figure 1). The \emph{structural network} $G_{S} = (V, E_{S})$ is formed by a set $V$ of $N$ nodes and a set $E_S$ of $L$ edges. The \emph{regulatory network} $G_{R} = (V, E_{S}, E_{R})$ is a signed bipartite network with one set of nodes given by $V$ (the nodes of the structural network), and another set of nodes given by $E_S$ (the edges of the structural network) connected by the regulatory interactions $E_{R}$ of cardinality $|E_{R}| = \hat{L}$.  The signed regulatory network can be encoded as an ${L}\times N$ matrix $K$ where $K_{\ell i}=1$  if node $i$ activates the structural  edge $\ell$,  $K_{\ell i}=-1$ if node $i$ inhibits the structural  edge $\ell$ and $K_{\ell i}=0$ otherwise.
If $K_{\ell i} = 1$, then the node $i$ is called a \emph{positive regulator} of the edge $\ell$, and if $K_{\ell i} = -1$, then the node $i$ is called a \emph{negative regulator} of the edge $\ell$. It is worth noting that node $i \in V$ cannot serve as both a positive and negative regulator for the same edge $\ell$ at the same time. However, node $i$ can act as a positive regulator for edge $\ell$ while simultaneously functioning as a negative regulator for a different edge $\ell^{\prime} \neq \ell$.

\section{The Triadic Perceptron Model (TPM)}
\label{model}
Here, we formulate a  model for node dynamics in a network with triadic interactions that we call Triadic Perceptron Model (TPM). The TPM  acts as a benchmark to validate the TRIM algorithm proposed here.
We assume that each node $i$ of the network is associated with a dynamical variable $X_i\in \mathbb{R}$, and  that the dynamical state of the entire network is encoded in the state vector $\vX=(X_1,X_2\ldots, X_N)^T$.
The topology of the structural network is encoded in the graph Laplacian matrix ${\bf L}$ with elements
\begin{align}
    L_{ij}
    &=
    \begin{cases}
        -a_{ij}J & \text{if}~~i \neq j, \\[0.25em]
        \sum_{k}a_{ik}J   & \text{if}~~i = j,
    \end{cases}\label{eqL2_v0}
\end{align}
where ${\bf a}$ is the adjacency matrix of the network of elements $a_{ij}$, and $J>0$ is a coupling constant.
In the absence of triadic interactions, we assume that the dynamics of the network is associated with a Gaussian process implemented as the Langevin equation
\bea
 \frac{d\vX}{dt}=-\frac{\delta \mathcal{H}}{\delta \vX}+\Gamma \bm{\eta}(t),
 \label{Lang1}
\eea
with the Hamiltonian
\bea
\mathcal{H}=\frac{1}{2}{\vX}^{\top}({\bf L}+\alpha {\bf I})\vX,
\eea
where $\Gamma>0$, $\alpha>0$,  and where ${\bm\eta}(t)$ indicates uncorrelated Gaussian noise with 
\bea
\Avg{\eta_i(t)}=0,\quad \Avg{\eta_i(t)\eta_j(t')}=\delta_{ij}\delta (t-t'),
\eea
for all $t$ and $t'$. The resulting Langevin dynamics are given by
\begin{align}\label{eq:node-state-dynamics}
    \frac{d\vX}{dt} = - (\mL + \alpha \mI) \vX + \Gamma \bm\eta({t}).
\end{align}
We remark that the Hamiltonian $\mathcal{H}$ has a minimum for $\vX={\bf 0}$, and its depth increases as the value of $\alpha$ increases. In a deterministic version of the model ($\Gamma = 0$), the effect of the structural interactions will not be revealed at stationarity. However, the Langevin dynamics with $\Gamma >0$ encode the topology on the network. Indeed, at equilibrium the correlation matrix $C_{ij}=\mathbb{E}((X_i-\mathbb{E}(X_i))(X_j-\mathbb{E}(X_j)))$ is given by 
\bea
C_{ij}=\frac{\Gamma^2}{2}[{\bf  L}+\alpha{\bf I}]^{-1}_{ij},
\label{Cij}
\eea
see Supplementary Information (SI) for details.
In other words, from the correlation matrix it is possible to infer the Laplacian, and hence the connectivity of the network.

We now introduce triadic interactions in the TPM. As explained earlier, a triadic interaction occurs when one or more nodes modulate the interaction between another two nodes. To incorporate triadic interactions into the network dynamics, we modify the definition of the Laplacian operator present in the Langevin equation. Namely, we consider the Langevin dynamics 
\bea\label{eq:triadic}
    \frac{d\vX}{dt} = - (\mL^{(\text{T})} + \alpha \mI) \vX  + \Gamma \bm\eta({t}),\label{eq:dyn}
\eea
obtained from Eq.(\ref{Lang1}) by substituting the graph Laplacian ${\bf L}$ with the 
\emph{triadic Laplacian} $\mL^{(\text{T})}$ whose elements are given by
\begin{align}
    L^{(\text{T})}_{ij}
    &=
    \begin{cases}
        - a_{ij}J_{ij} (\vX) & \text{if}~~i \neq j, \\[0.25em]
        \di\sum_{k=1}^{N} a_{ik}J_{ik} (\vX)  & \text{if}~~i = j.
    \end{cases}\label{eqL2}
\end{align}
Moreover, we assume that the coupling constants $J_{ij}(\vX)$ are determined by a perceptron-like model that considers all the regulatory nodes of the link $\ell = [i,j]$ and the sign of the regulatory interactions. Specifically, if $\sum_{k=1}^{N} K_{\ell k} X_{k}\geq \hat{T}$ then we set $J_{ij}=w^+$; if instead $\sum_{k=1}^{N} K_{\ell k} X_{k}<\hat{T}$ then we set $J_{ij}=w^{-}$, with $w_{+}, w_{-} \in \real_{+}$ and $w_{+} > w_{-}$. Thus, 
\bea
    J_{ij} (\vX)&=& w_{-}+  
        (w_{+}-w_{-}) \theta \left( \di\sum_{k=1}^{N} K_{\ell k} X_{k}- \hat{T} \right), 
\eea
where $\theta(\cdot)$ is the Heaviside function ($\theta(x)=1$ if $x\geq 0$ and $\theta(x)=0$ if $x<0$). 
Note that, in the presence of triadic interactions, the stochastic differential equation 
$(\ref{eq:triadic})$ is not associated with any Hamiltonian, and a stationary state of the dynamics is not guaranteed, making this dynamical process significantly more complex than the original Langevin dynamics given in Eq.(\ref{Lang1}).
The TPM is  related to  a recently proposed model that captures information propagation in multilayer networks \cite{nicoletti2023information}, but the TPM does not make use of a multilayer representation of the data. Moreover the TPM is  significantly different from models of higher-order interactions previously proposed in the context of consensus dynamics~\cite{neuhauser2022consensus,neuhauser2020multibody} or contagion dynamics~\cite{iacopini2019simplicial,de2020social}.
Indeed, in our framework, triadic interactions between continuous variables are not reducible to standard higher-order interactions because they involve the modulation of the interaction between a pair of nodes.  Moreover, this modulation of the interaction is not dependent on the properties of the interacting nodes and their immediate neighbors, as is the case in~\cite{neuhauser2022consensus,neuhauser2020multibody} or in the machine learning attention mechanism~\cite{
velivckovic2017graph}. On the contrary, the modulation of the interaction is determined by a third regulatory node (or a larger set of regulatory nodes) encoded in  the regulatory network.

{The TPM for continuous node dynamics in presence of triadic interactions is very general and comprehensively expresses the modulation of structural interactions by other nodes in the network. Therefore, the dynamics of TPM cannot be reduced to dynamics exclusively determined by pairwise interactions.} An important problem that then arises is whether such interactions can be mined from observational data. To address this issue, we will develop a new algorithm -- that we call the TRIM algorithm -- to identify triadic interactions from data, and we will test its performance on the data generated from the TPM model described above.
\begin{figure*}[!ht]
    \centering
\includegraphics[width=\linewidth,keepaspectratio]{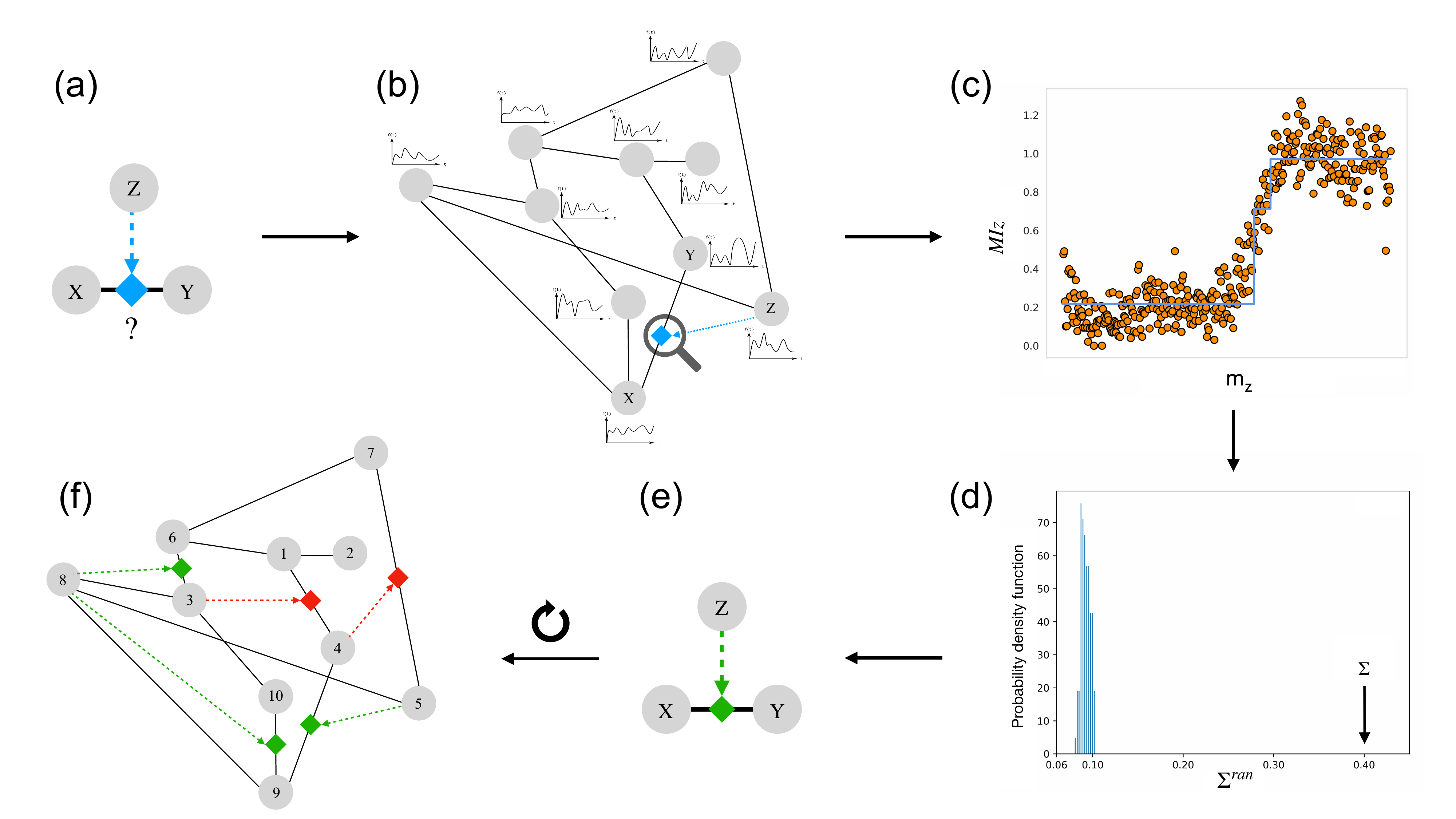}
    \caption{The TRIM algorithm identifies triples of nodes $X$, $Y$, and $Z$ involved in a putative triadic interaction, starting from the knowledge of the structural network and the dynamical variables associated with its nodes. For each putative triple of nodes involved in a triadic interaction (panel (a)) which belong to a network whose structure and dynamics is known (panel (b)), we study the functional behavior of the conditional mutual information $MIz$ (panel (c)), and assess the significance of the observed modulations of $MIz$ with respect to a null model (panel (d)). Given a predetermined confidence level, we can use these statistics to identify significant triadic interactions (panel (e)). This procedure can be extended to different triples of the network, thereby identifying the triadic interactions present in it (panel (f)).}
    \label{fig:2}
\end{figure*}

\section{Mining triadic interactions}
\subsection{The TRIM algorithm}
We propose the TRIM  algorithm (see Figure $\ref{fig:2}$) to mine triadic interactions among triples of nodes. To simplify the notation we will use the letters $X,Y,Z$ to indicate both nodes as well as their corresponding dynamical variables.

Given a structural edge between nodes $X$ and $Y$, our goal is to determine a confidence level for the existence of a triadic interaction involving an edge between node $X$ and node $Y$ with respect to a potential regulator node $Z$. Specifically, we aim to determine whether the node $Z$ regulates the edge between node $X$ and node $Y$, given the dynamical variables $X$, $Y$ and $Z$ associated with these nodes.  
To do so, given a time series associated with node $Z$, we first sort the $Z$-values, and define $P$ bins in terms of the quantiles of $z$, chosen in such a way that each bin $m_z$ comprises the same number of data points (ranging in our analyses from $30$ to $100$). We indicate with $z_m$ the quantile of $Z$ corresponding to the percentile $m/P$. Therefore, each bin $m_z$ indicates data in which $Z$ ranges in the interval $[z_m,z_{m+1})$. We indicate with $\mu(x|z_m),\mu(y|z_m)$ and $\mu(x,y|z_m)$ the probability density of the variables $X$, $Y$ and the joint probability density of the variables $X$ and $Y$ in each $m_z$ bin.

A triadic interaction is taken  to occur when the node $Z$ affects the strength of the interaction between the other two nodes $X$ and $Y$. Consequently, our starting point is to consider the mutual information between the dynamical variables $X$ and $Y$ conditional to the specific value of the dynamical variable $Z$. We thus consider the quantity
$MIz(m)=MI(X,Y|Z =z_m)$ defined as 
 \bea
    MIz(m)=\int dx \int dy \mu(x,y|z_m)\log\left(\frac{\mu(x,y|z_m)}{\mu(x|z_m)\mu(y|z_m)}\right).\nonumber
\eea
In order to estimate this quantity, we rely on non-parametric methods based on entropy estimation from $k$-nearest neighbours \cite{kozachenko1987sample,kraskov2004estimating,ross2014mutual} (see SI for details).
For each triple of nodes, we visualize the mutual information $MIz$ computed as a function of the $m/P$-th quantiles $z_m$ and fit this function with a decision tree comprising $r$ splits. 

In the absence of triadic interactions, we expect $MIz$ to be approximately constant as a function of the $m/P$-th quantiles $z_m$, while, in the presence of triadic interactions, we expect this quantity to vary significantly as a function of $z_m$. 
The discretised conditional mutual information CMI between $X$ and $Y$ conditioned on $Z$ can be written as 
\bea
CMI_{X,Y;Z}=\sum_{m=0}^{P-1}p(z_m)MIz(m)=\Avg{MIz},
\eea
where $p(z_m)=1/P$ indicates the probability that the $Z$ value falls in the $m_z$ bin.
This quantity indicates important information about the interaction between the nodes $X$ and $Y$ when combined with the information coming from the mutual information MI given by 
\bea
MI_{X,Y}=\int dx \int dy \mu(x,y)\log\left(\frac{\mu(x,y)}{\mu(x)\mu(y)}\right),\nonumber
\eea
where $\mu(x),\mu(y)$ are the probability density functions for $X$ and $Y$ and $\mu(x,y)$ is the joint probability density functions of the variables $X$ and $Y$.
 The conditional mutual information, however, is not sensitive to variations in $MIz$ and does not therefore provide the information needed to detect triadic interactions.
In order to overcome this limitation, we define the following two quantities that measure how much the mutual information between $X$ and $Y$ conditioned on $Z\in [z_m,z_{m+1})$ changes as $z_m$ varies.
Specifically we consider:
\begin{itemize}
\item[(1)]
 the standard deviation $\Sigma$ of $MIz$, defined as 
 \bea\Sigma = \sqrt{\sum_{m=0}^{P-1}p(z_m){\left[MIz(m)-\Avg{MIz}\right]^2}}; 
 \eea
 \item[(2)]
 the difference $T$ between the maximum and average value of $MIz$, given by  
 \bea
 T = \max_{m=0,\ldots, P-1}\left|MIz(m)-\Avg{MIz}\right|. 
 \eea
\end{itemize}
The quantities $\Sigma$ and $T$, collectively measure the strength of the triadic interaction under question and can thus  be used to mine triadic interactions in synthetic as well as  in real data. In order to assess the significance of the putative triadic interaction, we compare the observed values of these variables to the results obtained with given null models. 
In order to  determine if the observed values are significant with respect to a given null model, we compute the scores $\Theta_\Sigma$, $\Theta_T$, 
given by 
\bea
\Theta_\Sigma&=&\frac{\Sigma-\mathbb{E}(\Sigma^{\textup{ran}})}{\sqrt{\mathbb{E}((\Sigma^{\textup{ran}})^2)-(\mathbb{E}(\Sigma^{\textup{ran}}))^2}},\nonumber \\
\Theta_T&=&\frac{T-\mathbb{E}(T^{\textup{ran}})}{\sqrt{\mathbb{E}((T^{\textup{ran}})^2)-(\mathbb{E}(T^{\textup{ran}}))^2}},\nonumber \\
\eea
and the $p$-values
\bea
p_{\Sigma}=\mathbb{P}(\Sigma^{\textup{ran}}>\Sigma),\quad p_{T}=\mathbb{P}(T^{\textup{ran}}>T),
\eea
Note that if we consider $\mathcal{N}$ realizations of the null model, we cannot estimate probabilities smaller than $1/\mathcal{N}$. Therefore, if in our null model we observe no value of $\Sigma^{\textup{ran}}$ larger than the true data $\Sigma$, we set the conservative estimate $p_{\Sigma}=1/\mathcal{N}$. A similar procedure is applied also to $p_T$.

To assess this significance we consider two types of null models. The first is the randomization null model  obtained by shuffling the $Z$ values, to give $\mathcal{N}$ randomization of the data, i.e.~we use
surrogate data for testing~\cite{kurths1987attractor,theiler1992testing}. 
The second is the maximum likelihood Gaussian null model between the three nodes involved in the triple $X$,$Y$,$Z$. {{
Specifically, the Gaussian null model uses the mean and covariance of the timeseries of $X, Y$ and $Z$ to define a multivariate normal distribution from which samples are randomly drawn, thereby providing surrogate timeseries values for the considered triple.
}
We note that the use of these two null models allows us  to identify also non-monotonic relationships between $MIz$ and $z$, thereby going beyond underlying monotonic assumptions made elsewhere  \cite{Wang2020}. 
The first null model disrupts the temporal correlations between the timeseries of the node $Z$ and the timeseries the two nodes $X$ and $Y$ at the endpoints of the considered edges. Therefore this null model is robust with respect to the presence of possible outliers in the dataset. However, this first null model may overlook confounding network effects that affect correlations between the dynamical variables. The second null model more efficiently captures correlations between the dynamical state of the three considered nodes due to network effects but is more sensitive to the presence of outliers in the data.
To increase confidence, we therefore combined the insights coming from both these null models (see SI for details).

For each triple, the function $MIz(m)$ is fitted with a decision tree with two splits. In this way, three different intervals of values of $Z$ are identified, each corresponding to a distinct functional behavior of the correlation functions between the variables $X$ and $Y$. While our method in principle allows for more than two splits of the decision {tree}, for illustrative purposes we have chosen {two} splits since this is the minimum number of splits needed to capture non-trivial functional behavior in $MIz$, such as non-monotonicity. In practice this choice of two splits will also be the best choice when data is limited, such as the gene expression data we will analyze in the following section.

We also further characterize significant triples by examining their normalized entropic score function $S\in [0,1]$, which is used to characterize their corresponding functional behavior. {Specifically, th entropic score $S$ classifies the diversity of each of the joint distribution functions of $X$ and $Y$ conditioned on $Z$ for each interval obtained through the decision tree (see SI for details). 
}

As we will discuss below, the algorithm performs well on data obtained from the TPM. In this case, we also {observe that true triadic triples are characterized by a high entropic score $S$. On real data, the results obtained with the TRIM algorithm} using randomized surrogate data might neglect potentially important network effects, this shortcoming is mitigated by performing an additional {validation using the Gaussian null model and the entropic score $S$ (see SI for the full pipeline of TRIM)}. 
\begin{figure*}
    \centering
    \includegraphics[width=1.0\linewidth]{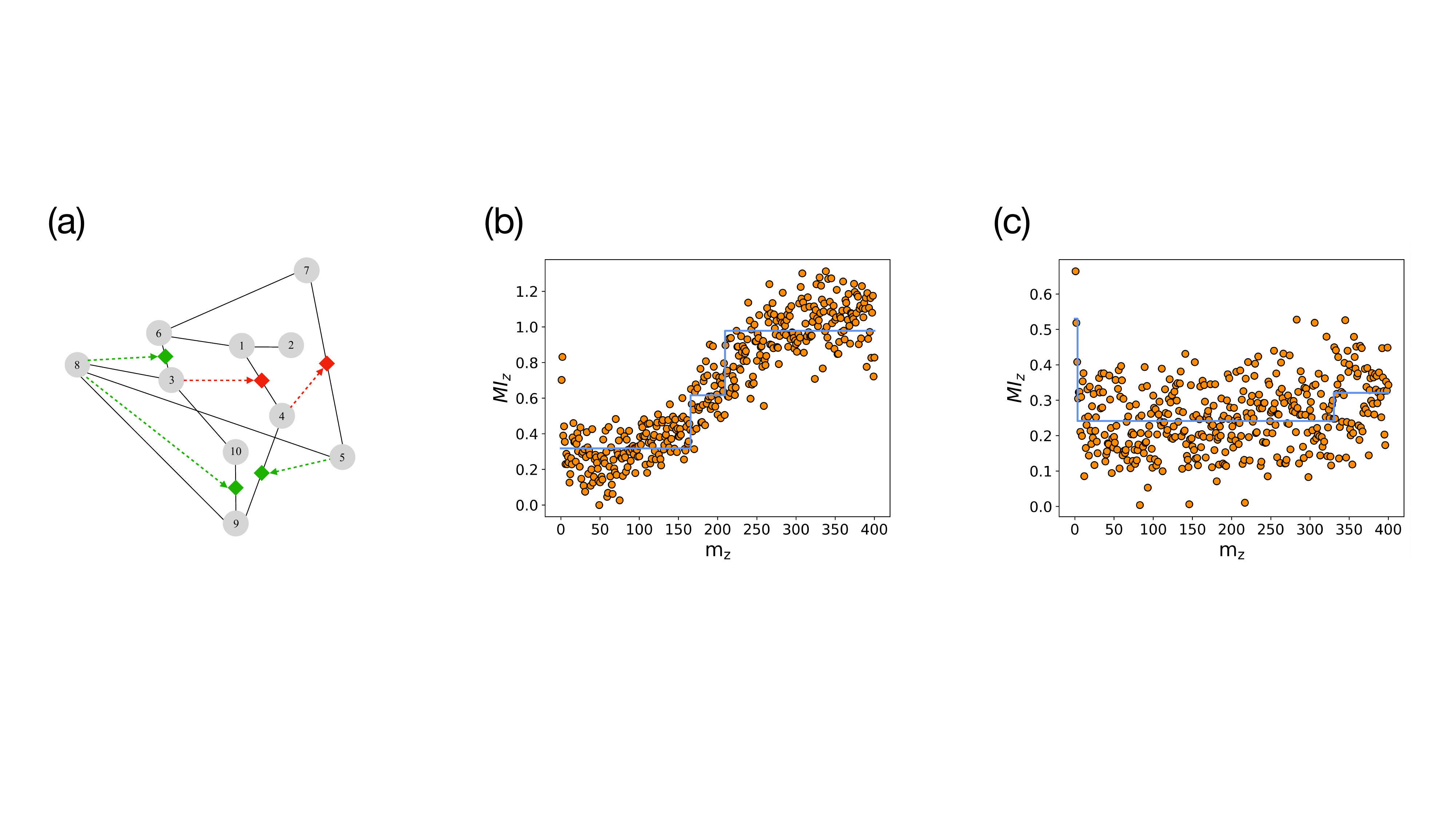}
    \caption{We consider a network with $N=10$ nodes, $L=12$ edges, and $\hat{L}=5$ triadic interactions (panel (a)). Panels (b) and (c) display the  effect of triadic interactions on the Mutual Information profile $MIz$. Panel (b) shows $MIz$ for the triple $[4,9,5]$ involved in a positive triadic interaction. Panel (c) shows $MIz$ for the triple $[1,2,6]$ that is not involved in a triadic interaction. In all panels simulations were run to $t_{\textup{max}}=4,000$ with a timestep of $dt=10^{-2}$. For the analysis we consider  $40,000$ time steps. The parameters of the model are: $\alpha=0.05,\hat{T}=10^{-3},  \Gamma=10^{-2}$,  $w^+=8, w^-=0.5$, number of bins {$P=400$}.}
    \label{fig:non-tri}
\end{figure*}

\begin{figure*}[!ht]
	\centering	\includegraphics[width=15cm,keepaspectratio]{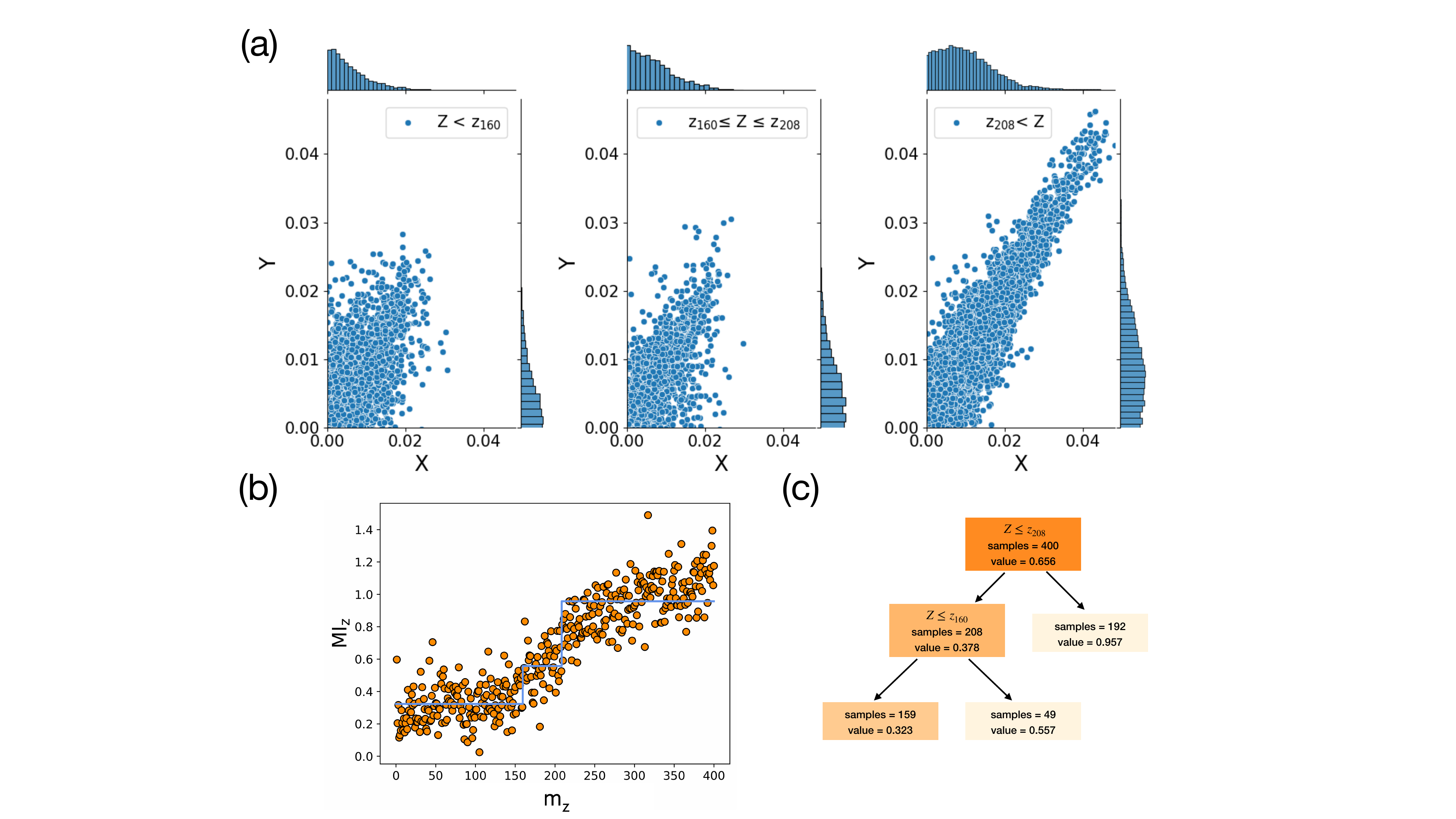}
 \caption{Representative results for triples of nodes involved in triadic interactions in the continuous model with triadic interactions. Results for the triple $[4,9,5]$ of the network in Figure 3 of the main text, which is triadic, are shown. The  joint distributions of variables $X$ and $Y$ conditional on the values of $Z$ are shown in panel (a). Panel (b) shows the behavior of $MIz$ as a function of the values of $z_m$, which clearly departs from the constant behavior expected in absence of triadic interactions. Panel (c) presents the decision tree for fitting the $MIz$ functional behavior and determining the range of values of $Z$ for which the most significant differences among the joint distributions of the variables $X$ and $Y$ conditioned on $Z$ are observed. The parameters used are the same as in Figure \ref{fig:non-tri}. }
 \label{fig:decisionT}
\end{figure*}

\begin{figure}[!ht]
	\centering
\includegraphics[width=0.99\linewidth,keepaspectratio]{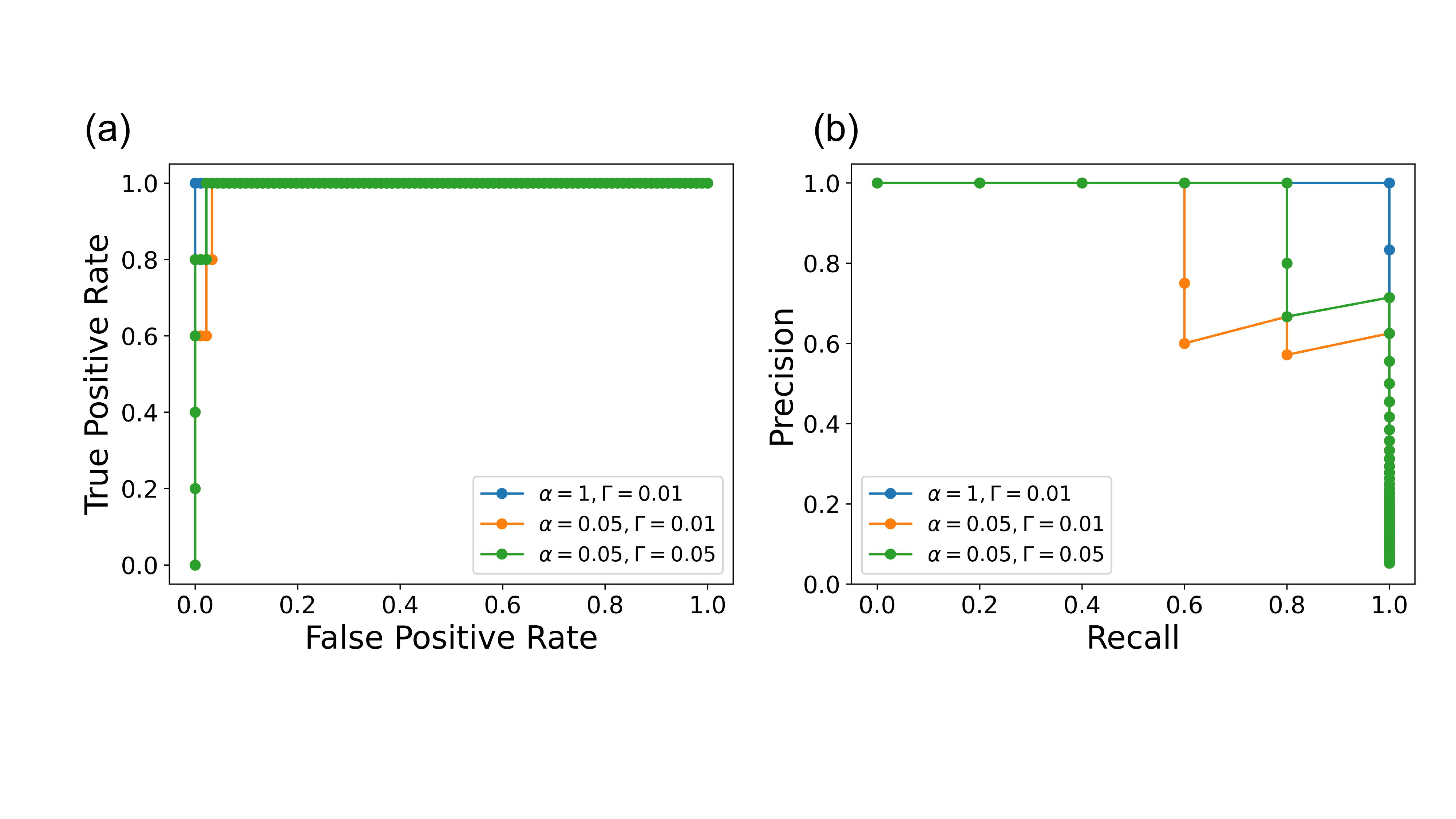}
 \caption{We consider the network in Figure \ref{fig:non-tri}(a). The time series obtained by integrating the stochastic dynamics of the proposed dynamical model for triadic interactions (Eq.(\ref{eq:dyn})) are analyzed with the TRIM algorithm. Panel (a) displays the Receiver Operating Characteristic curve (ROC curve) obtained by running TRIM with {$P=400$} bins and $\mathcal{N}=10^3$ realizations of the null model on these synthetic time series, using $\Theta_{\Sigma}$ to score for different parameters values indicated in the legend. Panel (b) displays the corresponding Precision-Recall curve (PR curve) obtained by running TRIM with the same parameters.
 The timeseries are simulated up to a maximum time $t_{max}=4000$ with a $dt=10^{-2}$. For the analysis, we consider $40,000$ time steps (see the SI for details). The parameter of the model are: $\hat{T}=10^{-3}$, $w^+=8$, $w^-=0.5$, and $\alpha$ and $\Gamma$ as indicated in the figure legend. }\label{fig:ROC}
\end{figure}

\subsection{Validation of the TRIM algorithm on the triadic perceptron model}

{In order to  discuss the phenomenology of the TPM  we first considered a representative network (see Figure $\ref{fig:non-tri}$) of $N=10$ nodes,  $L=12$ edges and $\hat{L}=5$ triadic interactions (each formed by a single node regulating a single edge) on top of which we consider the TPM proposed in Sec. \ref{model}}.

{We found that data obtained from the TPM on this network shows a strong dependence of $MIz(m_z)$ on $m_z$ for the triples of nodes involved in triadic interactions, with greater significance for smaller values of $\alpha$. Figure $\ref{fig:non-tri}$ shows the difference between the $MIz(m)$ profile of a triple that is involved in a  triadic interaction compared to a triple that is not, demonstrating how triadic interactions modulate the $MIz(m_z)$ profile.}

Moreover, Figure $\ref{fig:decisionT}$ shows, {for a given triadic interaction involving nodes X, Y and Z,} the joint distributions $\mu_\delta(X,Y)$ of $X$ and $Y$ for each interval $\delta$ of values $Z$ determined by the decision tree.  The results provide evidence of this interesting dynamical behavior of the triadic model in the case of a positive regulatory interaction. 
Note that the analysis of the form of the function $MIz(m_z)$ also allows us to distinguish between positive and negative regulatory interactions, which are associated with an increase or a decrease in $MIz$ for larger values of $m_z$ respectively.

In the Supplementary Figures S1-S2, we display further examples of the function $MIz$ for triadic triples. We observe the increased variability of the $MIz$ functional behavior as the parameter $\Gamma$ is raised, i.e. the noise increases. 

{These results confirm the main general principle on which the TRIM algorithm is based, i.e. that the conditional mutual information $MIz$ is modulated by triadic interactions. To make this observation precise, we examined the performance of the TRIM algorithm in mining triadic interactions from synthetic data. We first considered the network shown in Figure $\ref{fig:non-tri}$, and using the score $\Theta_{\Sigma}$, we evaluated the Receiver Operating Characteristic (ROC) curve and Precision Recall (PR) curve for different values of the dynamical parameters (see Figure $\ref{fig:ROC}$).
Both the ROC curve and the PR curve (which addresses the limitations of the ROC curve for imbalanced datasets) indicate that the TRIM algorithm performs well on data produced by the TPM, with a better performance for higher values of $\alpha$. }

For all parameter values, we noticed that false positives are more likely to involve short-range triples, i.e.~triples in which the regulator node $Z$ is close (in the structural network) to the end-points $X$ and $Y$ of the target edge.
\begin{figure*}[!ht]
	\centering
\includegraphics[width=15cm,keepaspectratio]{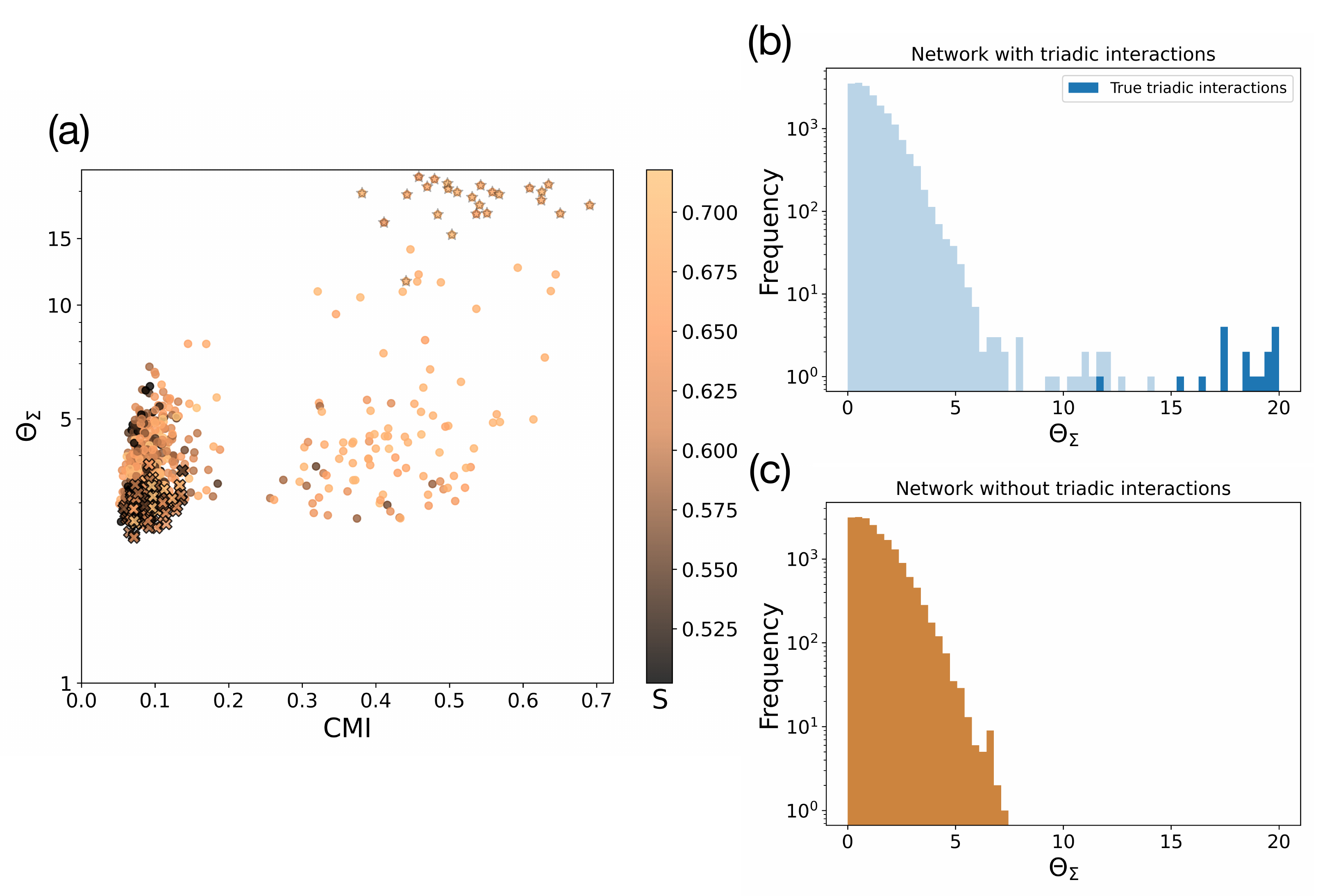}
	\caption{{Performance of the TRIM algorithm on a random network with triadic interactions. (a)} Each data point represents a given triple of nodes $X$, $Y$ and $Z$. The $y$-axis shows $\Theta_\Sigma$, while the $x$-axis shows the CMI between $X$ and $Y. $ The colour of each point corresponds to the value of $S$, which characterizes the entropic score of the triple. The synthetic data comes from a structural random {Erd\"os-Renyi } network with $100$ nodes, {and average degree $c=4$}, {to which $25$ triadic interactions between random edges and random nodes have been added. For the modelling of the network we used $\alpha = 0.06, \Gamma=1.4\times 10^{-2}, t_{max}=1500, w_+=18,w_-=0.2$ and for the analysis with TRIM we looked at $3000$ data points and $P=100$ bins.} We display top 5 triples for each edge according to $\Theta_\Sigma$ that are below our $p$ threshold for the randomization null model. The Triples below are represented in the scatter plot and they all display an entropic score $S>0.5$. Stars are the true triadic triples which are characterized by high $\Theta_{\Sigma}$. {Crosses are the triples that can be excluded by performing TRIM with the Gaussian Null model. }{(b) Histogram of the $\Theta_\Sigma$-values for all the triples of the network (in light blue), and for the triples corresponding to the $25$ true triadic interactions only (in dark blue). (c) Histogram of the $\Theta_\Sigma$-values observed in a network of the same topology and with the same dynamical parameters for which all the triadic interactions have been removed (orange).}}
 \label{fig:syn}
\end{figure*}
\begin{figure*}[!htb!]
    \centering
    \includegraphics[width=15cm,keepaspectratio]{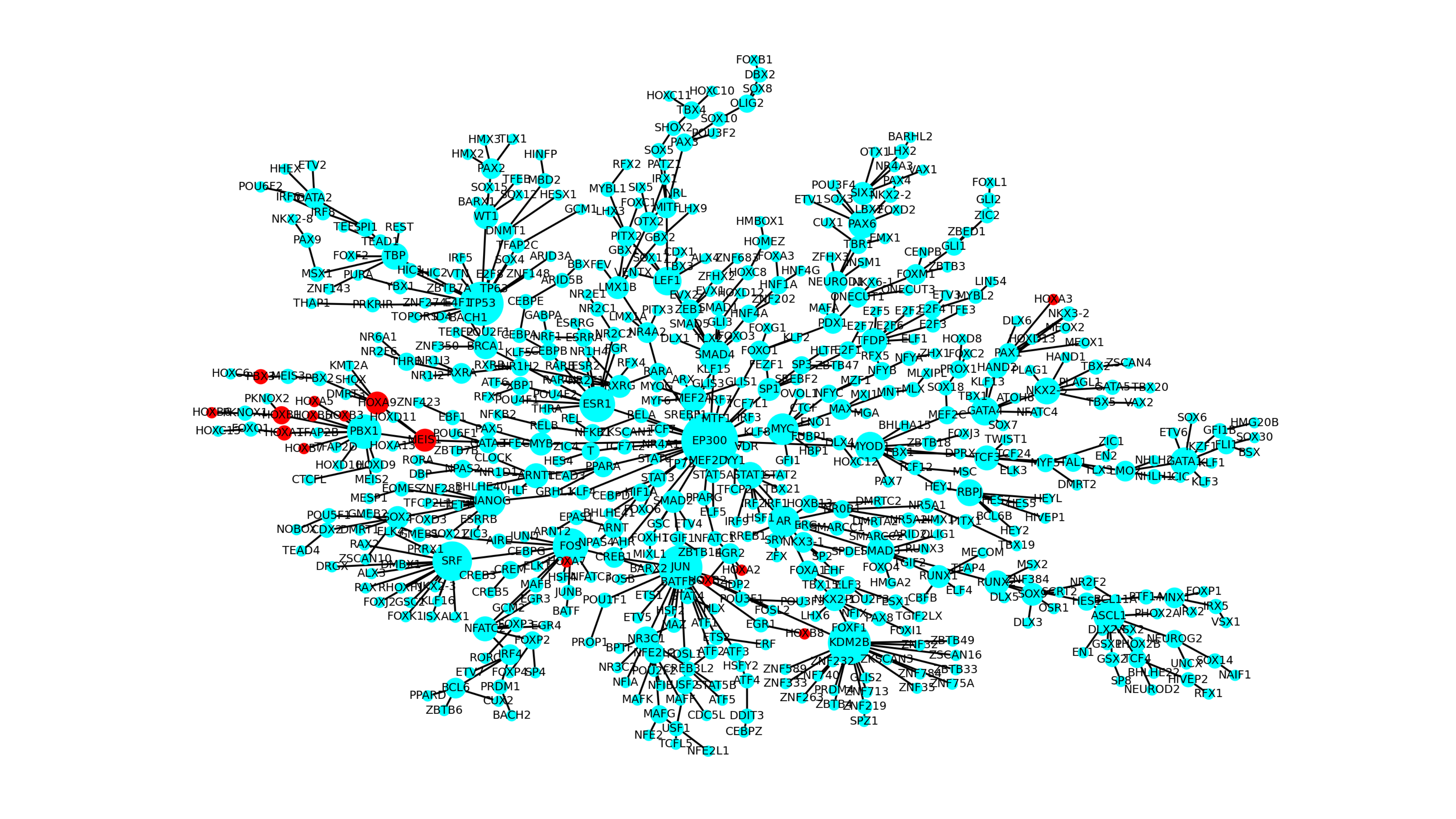}
    \caption{Maximal Spanning Tree of the relevant genes from the gene expression data. Edges and their edge weights were obtained from the Protein-Protein interaction network. Red-coloured nodes are nodes with biological significance, that is that play a critical role in AML~\cite{Alharbi2013,Guo2017,Xiang2022}. Node size is proportional to the node degree.}
    \label{fig:MST}
\end{figure*}
\begin{figure*}[!htb!]
	\centering
\includegraphics[width=0.99\linewidth,keepaspectratio]{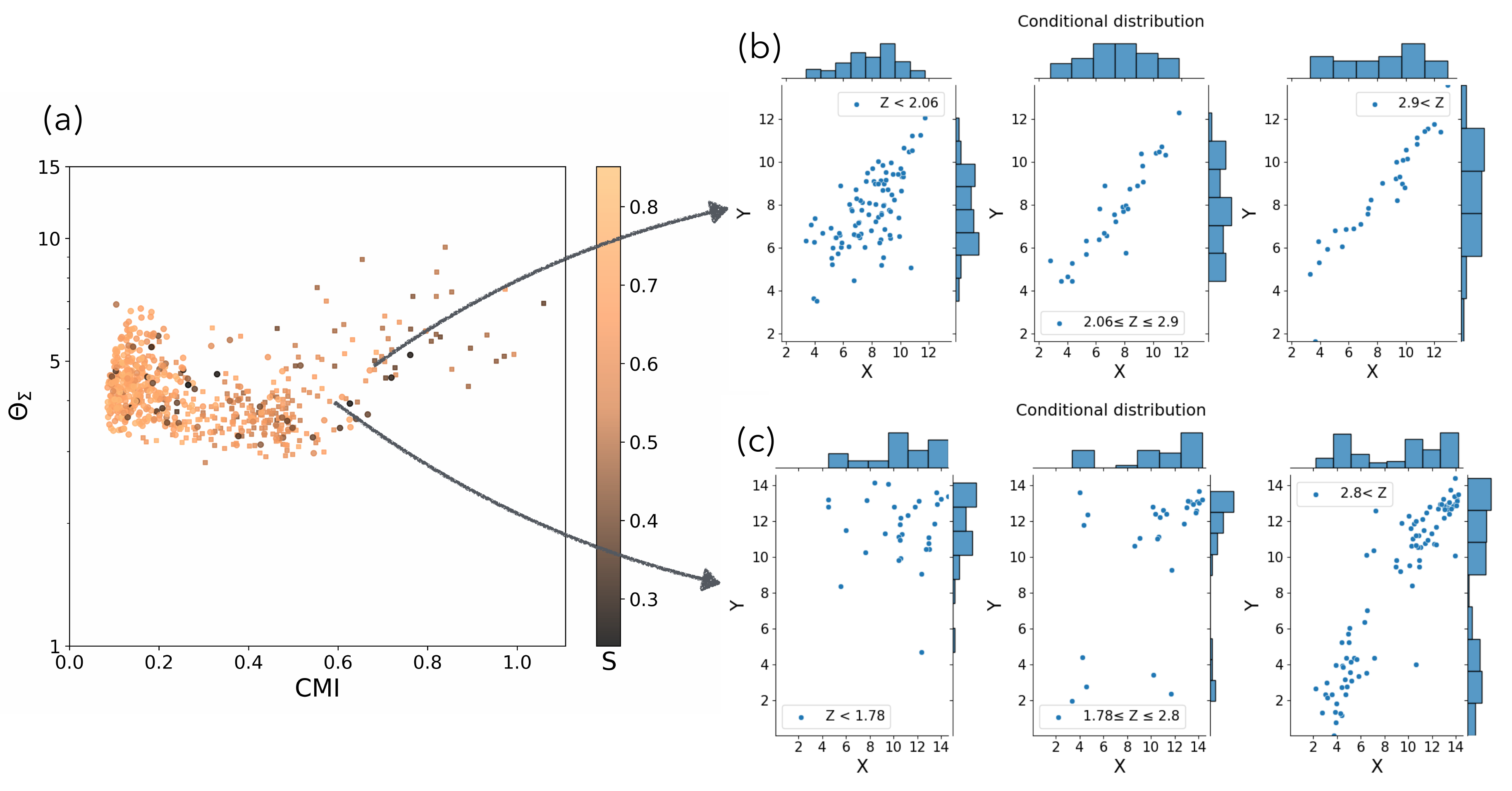}
	\caption{Application of the TRIM algorithm to gene expression data.
   Panel  (a)  shows the results of TRIM for the significant triples in the AML dataset. The scatter plot shows $\Theta_\Sigma$ ($y$-axis) versus the CMI ($x$-axis). The colour of each point corresponds to the value of its entropic score $S$. Here we display only those triples with $p$-value 0.001 or less in the randomization null model and that have not been excluded by the Gaussian null model (for details about these triples see SI). Circles are triples whose links all appear in the minimum spanning tree, and squares indicate triples involving genes with biological relevance. Panels (b)-(c) display the conditional distributions  for two example triples: both are identified by the TRIM algorithm with high significance, suggesting a meaningful biological association. Panel  (b) shows the triple $X$ = \textit{GATA1}, $Y$ = \textit{KLF1}, $Z$ = \textit{ETV1}. According to the randomized surrogate null model, this triadic interaction has $p_{\Sigma}$-value 0.001, $\Theta_\Sigma$ = 4.75, $\Sigma$ = 0.44, $S = 0.64$; panel (c) shows  the results for the triple $X$ = \textit{HOXB3}, $Y$ = \textit{MEIS1}, $Z$ = \textit{GLIS3} involving two biologically relevant genes. According to the randomized surrogate null model, it has $\Theta_\Sigma$ = 3.98, $p_\Sigma$ = 0.001, $\Sigma$ = 0.38, $S = 0.60$.}
   
 \label{fig:7}
\end{figure*}
These results indicate that the TRIM algorithm is effective in identifying triadic interactions in a small network generated using the TPM. To examine the scalability of this methodology we also tested the TRIM algorithm on a much larger model network. To this end, we consider a random {Erd\"os-Renyi }network of $100$ nodes, {and average degree $c=4$ to which we added $25$ random triadic interactions (i.e., between randomly chosen nodes to randomly chosen edges), imposing the condition that each edge is at most regulated by a single node for simplicity.} The results of this analysis are shown in Figure $6(a)$, in which we provide statistics for all possible node triples in the network (the majority of which are not triadic interactions). For each edge, we retained only the $5$ most significant triples according to $\Theta_{\Sigma}$. By conditioning on the value of  third node, for each of these connected nodes we also record the conditional mutual information CMI.  In Figure 6, each considered triple corresponds to a point, colour coded according to the value of $S$. Stars indicate triples that are involved in a triadic interactions  (see SI for details).
True triadic interactions are found for triples with high $\Theta_{\Sigma}$ while CMI span between high and intermediate values.  {This result confirms the  very good  performance of the TRIM algorithm on the data coming form the TPM.}

{To test the statistical robustness of the TRIM algorithm we also conducted the same analysis (i.e., on the same structural network with the same dynamical parameters) in which all triadic interactions were removed. The results of this analysis are shown in Figure $\ref{fig:syn}$(c). In this case, and as expected, the TRIM algorithm did not identify any statistically significant triadic interactions. This analysis indicates that the TRIM algorithm is able to identify true triadic interactions with a low false positive discovery rate (compare Figures $\ref{fig:syn}$(b)-(c)).}

\section{Detecting triadic interactions in gene-expression data}

Searching for triadic interactions in gene-expression is a problem of major interest in biology. For instance, understanding the extent to which a modulator promotes or inhibits the interplay between a transcription factor and its target gene is crucial for deciphering gene regulation mechanisms \cite{Wang2009}. In order to address this question with our method, we considered a gene-expression dataset associated with Acute Myeloid Leukemia (AML), extracted from the Grand Gene Regulatory Network Database~\cite{Gene_expression_data,BenGuebila2022}. 

Exhaustive mining of all potential triadic interactions from every putative triple of nodes in the AML dataset is computationally very demanding (it would require testing of $>260M$ triples) and likely, due to the sheer number of triples being tested, to result in false positives and/or interactions of less biological importance. Moreover, such a brute-force approach would not account for other sources of important biological information, such as putative interactions derived from other experimental sources. To account for such information, we therefore focused our analysis on edges between nodes associated with known biophysical interactions, as identified in the human Protein-Protein Interaction network (PPI) ~\cite{Gene_expression_data}.  
To do this, we considered the connected subgraph of the human PPI network that contains all the genes/proteins included in the AML gene expression data and their associated edges This network, which contains $622$ nodes and $42,511$ edges, formed the structural network for our analysis \cite{grigoriev2001}.
To start, we focused on triples involving genes known to be associated with AML, in which the end-points $X$ and $Y$ of the target are directly connected in the PPI network (see SI for details). 
 
We then selected additional triples according to their positions in the PPI network's Maximum Spanning Tree (MST), which only includes 621 edges (see Figure \ref{fig:MST}).  In order to focus on triples for which network effects are likely to be less pronounced, for each edge in the MST, connecting gene/protein $X$ with gene/protein $Y$ we considered all genes $Z$ within a distance of 4 from both the $X$ and $Y$ as candidate regulatory nodes, i.e. the third node in the triple (see SI for details). For each considered triple of genes we assessed its significance using $\Theta_\Sigma$ as the significance score, with $P=5$ bins, using $\mathcal{N}=5\times 10^3$ realizations of the randomization null model (very similar results were obtained using $\Theta_T$ as the significance score, {see Supplementary Figures S3-S4} for a comparison). 

Figure \ref{fig:7} shows the results of the TRIM algorithm for those triples with $p_\Sigma<0.001$. 
 Note that for  each selected edge only the top $5$  triples ranked according to the  $\Theta_\Sigma$ score are depicted. Squares indicate triples chosen from biologically relevant genes for AML. The triples deemed insignificant according to the Gaussian null model, are not shown here. The interested reader can see their visualization in Figure $S6$ of the SI.}
Figure \ref{fig:7} provide examples of conditional distributions of two illustrative triples that rank high in TRIM. Both triples show evidence of a triadic interaction: the triple in panel (b) is a member of the MST, while (c) is an example of a triple chosen from known biologically relevant genes for AML. Further example triples are shown in the SI. Interestingly, among  the significant triples, we detected also triples in which the modulation of the mutual information is non-monotonic (see SI for details). 

Many of the genes involved in the $50$ highest ranking triples have already been linked to AML in the literature (see SI for Table S3 for a list of highly significant triples and Table S4 with links to the literature associating the involved genes with AML).
In total, $84\%$ of the top 50 Triples include at least one gene that has a known association with AML.

\section{Conclusions}
This work provides a comprehensive information theory-based framework to model and mine triadic interactions directly from dynamic observations. The TPM  we propose demonstrates that the presence of a triadic interaction leads to systematic variations in the  mutual information between the two end nodes of the edge involved ($X$ and $Y$). Via this model we have shown that to detect triadic interactions it is necessary  to  go beyond standard pairwise measures, such as the mutual information. Importantly, standard higher-order statistical measures, such as the conditional mutual information, which accounts for the average effect of the third regulatory node $Z$ on the mutual information between the target nodes $X$ and $Y$ are also insufficient to identify triadic interactions. 
Our proposed approach, implemented in the TRIM algorithm, mines triadic interactions by identifying statistically significant variations in the mutual information between the two linked nodes conditioned on the third regulator node.

To demonstrate the efficacy of this algorithm we have tested and validated it on a new dynamical model (that we denote the TPM) and shown how it can identify triadic interactions in randomly generated triadic interaction networks. We also used it to mine  putative triadic interactions from gene expression data, and connect the putative interactions with meaningful biology. 

From the network theory point of view, this work opens new perspectives in the active field of modelling and inference of higher-order interactions and can be extended in many different directions, for instance by exploring the effect of triadic interactions on the dynamical state of nodes associated with discrete variables {or including time delays in the regulation}.  From the biological point of view, our results  may inspire further information-theoretic approaches to genetic regulatory network inference. Investigating the extent to which triadic interactions are tissue-specific, and if certain regulatory patterns are conserved across different tissues, could yield valuable insights.
Our proposed approach could also be used to mine triadic interactions in other domains, such as finance or climate, where triadic interactions also have a significant role.

\section*{Code availability}
The Python package TRIM is available on GitHub at the following link: \href{https://github.com/anthbapt/TRIM}{https://github.com/anthbapt/TRIM}

\section*{Acknowledgments}
This research utilized Queen Mary’s Apocrita HPC facility, supported by QMUL Research-IT, https://doi.org/10.5281/zenodo.438045. This work was sponsored by the Turing-Roche Strategic Partnership.

\onecolumngrid

\appendix
\newpage
\newpage

\renewcommand\theequation{{S-\arabic{equation}}}

\renewcommand\thetable{{S-\Roman{table}}}

\renewcommand\thefigure{{S-\arabic{figure}}}

\setcounter{equation}{0}

\setcounter{figure}{0}

\setcounter{section}{0}

\begin{center}

\Large{\bf SUPPLEMENTAL MATERIAL}

\end{center}

\section{Supplementary information on the TRIM algorithm}

\subsection{ Estimation of the Mutual Information}
{The Mutual Information  $MIz(m)$ between variables $X$ and $Y$ conditioned on the value $z_m$ of the variable $Z$ (putative regulator node of the triadic interaction) is calculated using k-nearest-neighbour entropy estimation to be able to estimate it only by using the time series values. \cite{kozachenko1987sample,kraskov2004estimating,ross2014mutual}, $MIz(m)$ is defined as: 
\bea
    MIz(m)=\int dx \int dy \mu(x,y|z_m)\log\left(\frac{\mu(x,y|z_m)}{\mu(x|z_m)\mu(y|z_m)}\right).
\eea
By assuming some metric, neighbours of a chosen point can be ranked by distance. The idea is to estimate from the average distance to the nearest neighbour, averaging over all. This can be used to estimate the logarithms of the entropies using the Kozachenko-Leonenko-Estimator \cite{kozachenko1987sample}.}

\subsection{Entropic score $S$ for significant triples }
In order to identify and classify the significant triples $[X,Y,Z]$ involving node $X$ and $Y$ whose interaction is modulated by node $Z$,   we introduce an entropic score function $S$ which  characterizes how diverse the conditional joint distributions ${\mu_\delta}(X,Y)$ of $X$ and $Y$ conditioned on $Z$ in each of the obtained intervals $\delta\in \{1,2,3\}$ are. Dividing the plane $X,Y$ in ${P}^2$ squares $(i,j)$ (by binning $X$ and $Y$ in ${P}$ bins each) with $n^{(\delta)}_{ij}$ data points,  we can calculate the participation ratio $Y_2^{(\delta)}$ \cite{barthelemy2003spatial,derrida1987statistical}
\bea
Y_2^{(\delta)}=\sum_{i=1}^{{P}}\sum_{j=1}^{{P}} \left(\frac{n_{ij}^{(\delta)}}{\mathcal{N}^{(\delta)}}\right)^2,
\eea
where $\mathcal{N}^{(\delta)}=\sum_{i=1}^{{P}}\sum_{j=1}^{{P}} {n_{ij}^{(\delta)}}$.
The inverse of the partition function is known to measure the effective number of square bins in which the distribution is localized.
We can then introduce the normalized entropic score $S$ as 
\bea
S=-{\frac{1}{3 \ln P^2}}\sum_{\delta=1}^3 \ln Y_2^{(\delta)}.
\eea
The entropy $S$ is low if  all the conditional distributions ${\mu_\delta}(X,Y)$ are very localized while it acquires large values if all the conditional distributions are delocalized.
{{We adopt a threshold $S=0.5$ in order to retain triples with $S>0.5$ indicating that in average the conditional distributions associated to these triples have more than $\sqrt{P^2}$ significantly populated bins.}}\\
\subsection{Pipeline of TRIM}

{In order to select for the significant triples, we combine information coming from the two considered null models (the surrogate randomised data and the maximum likelihood Gaussian model) and the entropic score $S$. 
First we select the set of triples of interest, and we choose which observable to consider, either $\Sigma$ or $T$. Note that we have found that the results obtained considering either one of the two observables are highly correlated (see section on gene-expression results).}

{With respect to the randomized null model, the Gaussian null model more efficiently captures correlations between the dynamical state of the three considered nodes due to network effects. However, the second null model is more sensitive to the presence of outliers in the data but it is still focusing only on the three nodes in the triple and neglects more collective network effects. In order to further reduce the set of relevant triples, we screen out triples with low entropic score.
Thus,  we define the following TRIM pipeline:
\begin{itemize}
\item[(a)] For each link of interest between variables X and Y  consider all the triples X,Y, Z where Z is any given possible regulator node (any node different from X and Y) and calculate $\Theta_{\Sigma}$ or $\Theta_T$ together with their corresponding $p$-values. Select only triples with p-values smaller than a threshold (typically taken $5\times 10^{-3}$ or $1\times 10^{-3}$). 
\item[(b)] 
Screen out all the triples which have a high p-value according to the second (Gaussian) null model.  Here a high p-value indicates a p-value higher than a threshold typically taken the same as in point (a)).
\item[(c)] Rank the remaining triples that rank high according to the $\Theta$ score calculated according to the randomized surrogate data. For each link between node X and Y select only the top 5 triples [X,Y,Z].
\item[(d)] 
From the remaining triples
 screen out all the triples which have a low entropic score $S$ (for instance $S<0.5$).
\end{itemize}
}

\section{Supplementary information on the Triadic  Perceptron Model (TPM)}
\subsection{\bf Derivation of Eq.(6) of the main text}
{Let us consider the stochastic dynamics Eq.(5) in the main text, driving the TPM in the absence of triadic interactions, i.e.
\bea
\label{eq:node_state_dynamics}
    \frac{d{\bf X}}{dt} = - ({\bf L} + \alpha {\bf I}) {\bf X} + \Gamma \bm\eta({t})
    \label{st}
\eea
where $\mL$ is the graph Laplacian of the networks $\alpha,\Gamma\in \mathbb{R}^+$ are two real parameters of the model, and $\bm\eta$ is a Gaussian white noise  with
\bea
\Avg{\eta_i(t)}=0,\quad \Avg{\eta_i(t)\eta_j(t')}=\delta_{ij}\delta (t-t').
\eea
Here we want to derive Eq.(6) of the main text expressing the correlation matrix $C_{ij}=\mathbb{E}((X_i-\mathbb{E}(X_i))(X_j-\mathbb{E}(X_j)))$ at equilibrium, i.e.
\bea
C_{ij}=\frac{\Gamma^2}{2}[{\bf  L}+\alpha{\bf I}]^{-1}_{ij}.
\label{Cij}
\eea}

{The derivation of this results is a straightforward outcome of the solution of the Fokker-Planck equation associated to  the stochastic dynamics $(\ref{st})$.
Indeed the Fokker-Planck equation for the probability density function $P({\bf X})$ of observing ${\bf X}$ reads 
\bea
\frac{\partial P({\bf X})}{\partial t}=-\sum_{i=1}^N\frac{\partial}{\partial X_i} [({\bf L} + \alpha {\bf I}){\bf X}]_{i}P({\bf X})+\frac{\Gamma^2}{2}\sum_{i,j=1}^N\frac{\partial^2}{\partial^2 X_i}P({\bf X}).
\eea
At stationarity, by imposing ${\partial P({\bf X})}/{\partial t}=0$ and solving for $P({\bf X})$, we obtain a multivariate Gaussian distribution 
\bea
P({\bf X})=\mathcal{C}\exp\left[-\frac{1}{2}{\bf X}^{\top}{\bf C}^{-1}{\bf X}\right],
\eea
where $\mathcal{C}$ is the normalization constant, and ${\bf C}$ indicates the covariance matrix is given by Eq.(\ref{Cij}).}

\section{Supplementary Information on the TPM dynamical behaviour}
{Here we provide supplementary information on the dynamics of the TPM (Eq. \ref{eq:node_state_dynamics}) on the network with $N=10$ nodes, $L=12$ edges and $\hat{L}=5$ regulatory interactions shown in Figure 4 of the main text. We consider the time series obtained by integrating the stochastic dynamics of the proposed dynamical model for triadic interactions within continuous variables. The time series is simulated up to a maximum time $t_{\textup{max}}=4000$ with a  $dt=10^{-2}$ leading to $4\times 10^{5}$ data points.
For our analysis we consider the last $200,000$ data points sampled every fifth point leading to   $40,000$ time steps. 
In the Supplementary Figures \ref{S1} and \ref{S2} we report the analysis conducted on the  triadic triple [4,9,5] already considered in Figure 4 but  for different parameter values: for Figure \ref{S1} we have $\alpha=0.01, \Gamma=10^{-2}$, for Figure  \ref{S2} we have $\alpha=0.05, \Gamma=5\times 10^{-2}$. The other parameter values are:  $\hat{T}=10^{-3}, w^+=8, w^-=0.5$, number of bins {$P=400$}.}

\begin{figure}[!ht]
	\centering
	\includegraphics[width=15cm,keepaspectratio]{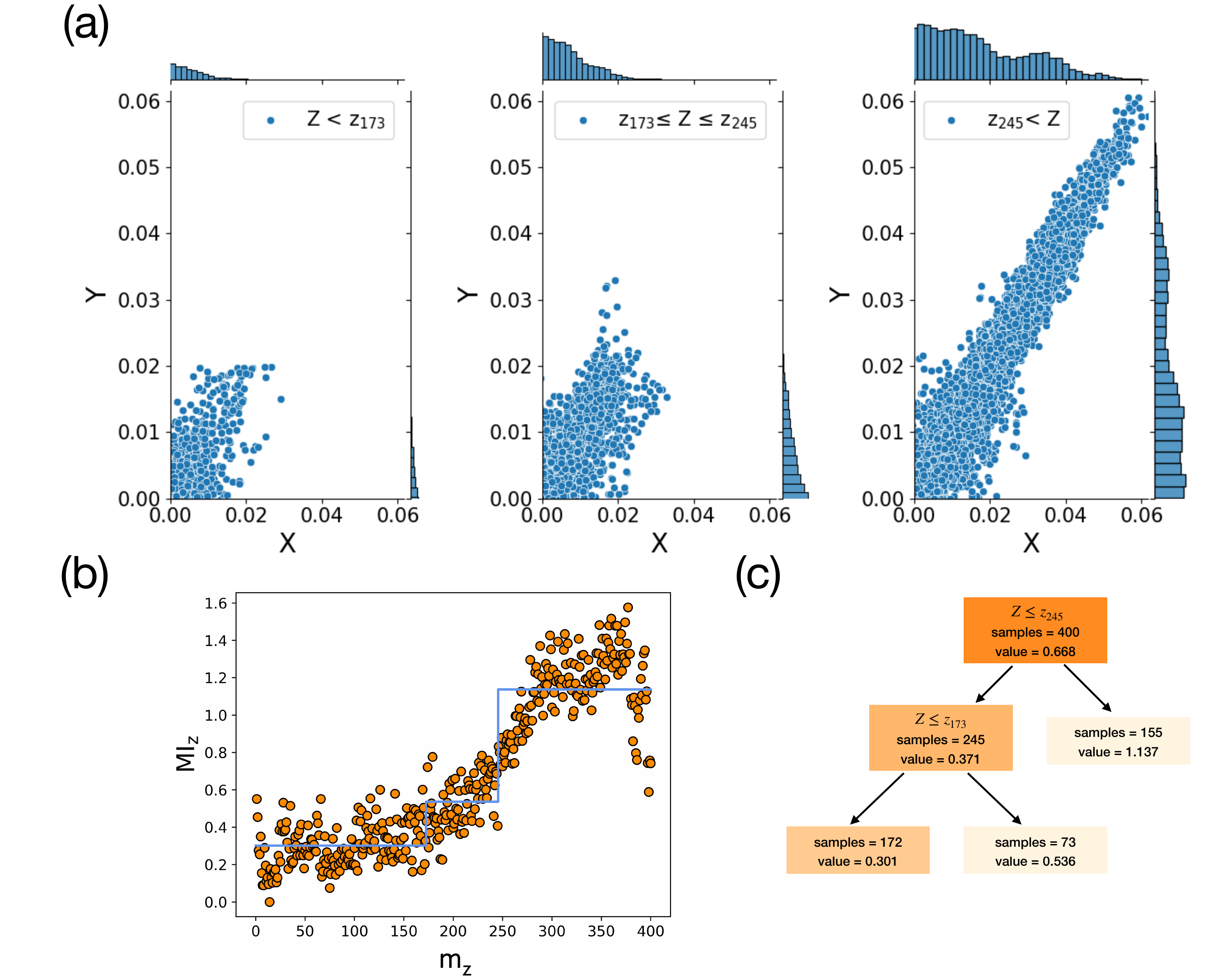}
	\caption{Exemplary results obtained for a triple of nodes involved in a triadic interaction in the continuous model with triadic interactions. The  joint distributions of variables $X$ and $Y$ conditional on the values of $Z$ is shown in panel (a). Panel (b) displays the functional behaviour of $MIz$ as a function of the values of $z_m$ which clearly departs from the constant behaviour expected in absence of triadic interactions. Panel (c) presents the decision tree for fitting the $MIz$ functional behaviour and determining the range of values of $Z$ for which the most significant differences among the joint distributions of the variables $X$ and $Y$ conditional to $Z$ are observed.  The time series is simulated up to maximum time $t_{\textup{max}}=4 000$ with $dt=10^{-2}$. For the analysis we consider  $40,000$ time steps.  The parameters of the model are: $\alpha=0.01,\hat{T}=10^{-3},$ $\Gamma=10^{-2}$,  $w^+=8, w^-=0.5$, number of bins {$P=400$}. The analysis is done for the triple [4,9,5], of the network in Figure 3 of the main text, which is triadic.}
 \label{S1}
\end{figure}

\begin{figure}[!ht]
	\centering
	\includegraphics[width=15cm,keepaspectratio]{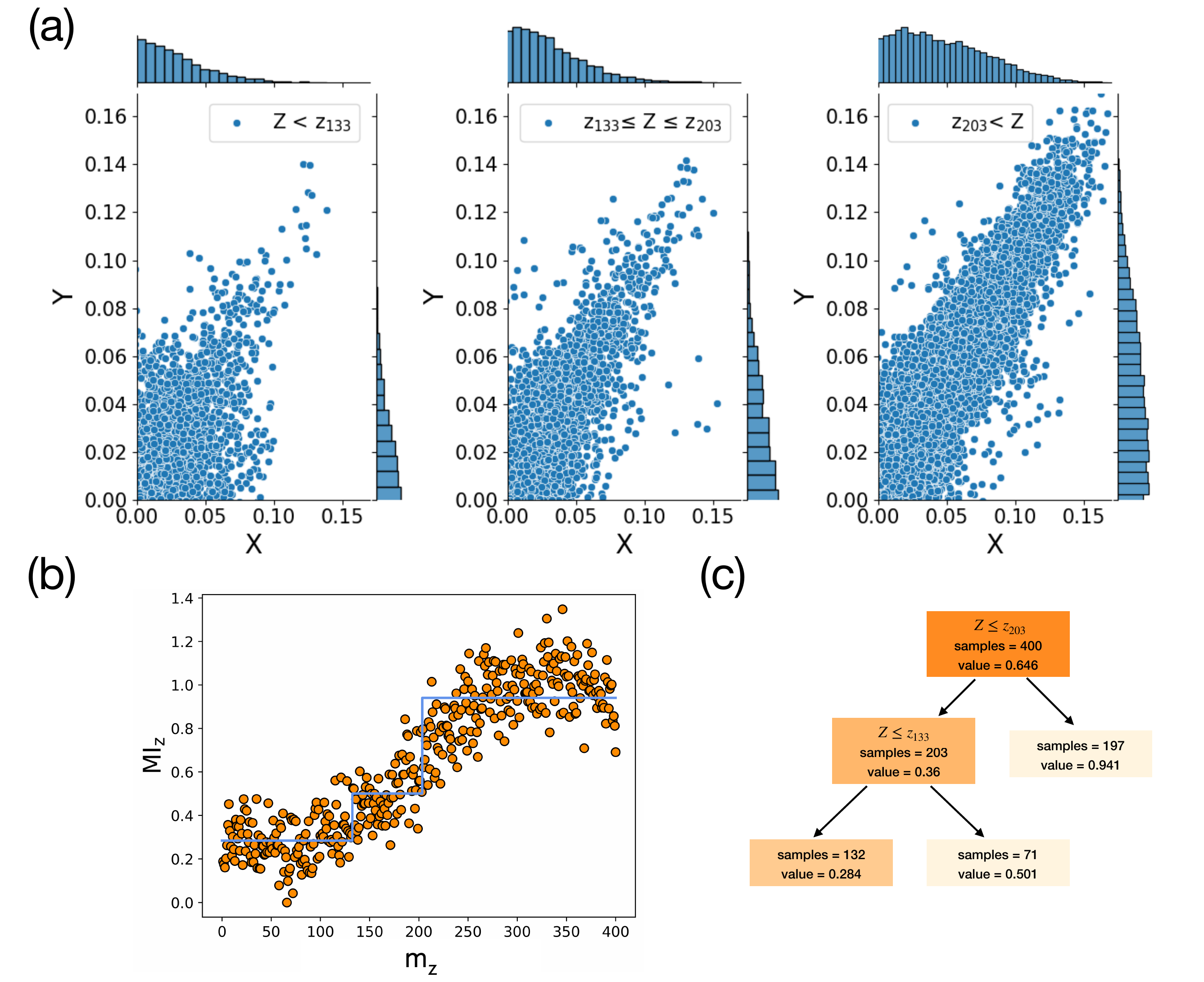}
	\caption{ Exemplary results obtained for a triple of nodes involved in a triadic interaction in the continuous model with triadic interactions. The  joint distributions of variables $X$ and $Y$ conditional on the values of $Z$ is shown in panel (a). Panel (b) displays the functional behaviour of $MIz$ as a function of the values of $z_m$ which clearly departs from the constant behaviour expected in absence of triadic interactions. Panel (c) presents the decision tree for fitting the $MIz$ functional behaviour and determining the range of values of $Z$ for which the most significant differences among the joint distributions of the variables $X$ and $Y$ conditional to $Z$ are observed. The time series is simulated up to maximum time $t_{\textup{max}}=4000$ with $dt=10^{-2}$. For the analysis we consider  $40,000$ time steps.  The parameters of the model are: $\alpha=0.05,\hat{T}=10^{-3},$   $\Gamma=5\times 10^{-2}$,  $w^+=8, w^-=0.5$, number of bins {$P=400$}.The analysis is done for the triple [4,9,5], of the network in Figure 3 of the main text, which is triadic. }
 \label{S2}
\end{figure}
\section{Supplementary information on analysis of gene-expression dataset}
\subsection{Methods}
We choose the gene-expression associated with Acute Myeloid Leukemia (AML) extracted from the Grand Gene Regulatory Network Database~\cite{Gene_expression_data,BenGuebila2022} for our TRIM analysis. Testing all possible combinations of triples of genes is computationally too demanding. To still make the analysis rigorous, we  focus on the most important genes and the genes we suspect are most likely to be involved in triadic interactions. Hence, we utilize the Protein-Protein Interaction (PPI) network associated with AML, which captures digenic processes, to choose edges forming the triples in our study. This selection is motivated by studies indicating that for gene expression data, most of the genes involved in trigenic processes are also involved in digenic processes \cite{Kuzmin2018}. We consider all the genes in the AML datasets that are connected in the PPI network. Each edge in the PPI network has an assigned edge weight. This allows us to calculate the maximum spanning tree (MST) (Figure 7). 
The MST has diameter $16$ and includes $622$ nodes and $621$ edges. `EP300' is the node with the largest node degree. To further reduce the number of triples that are being analysed, only triples where the distance of potential regulator node and edge are at least $4$ in the maximal spanning tree are being chosen. This is to reduce the effect of the triple being a triangle where all nodes are connected by an edge and to filter out dependencies from nodes of the triple being connected trough only a small number of edges.

Using this strategy $343,194$ triples are being analysed. To further decrease computational time, the number of realisations that the $\Sigma$ or $T$ values are being compared to is adjusted based on the confidence interval for a triadic interaction of the triple being investigated. First $10$ realisations are run. If the $p_\Sigma$-value is bigger than $0.3$, the triple is being categorised as non-triadic. In the other case, 100 realisations will be run and if the $p_\Sigma$-value of the new result is smaller than 0.05, 1000 realisations will be run. This is to quickly exclude any triplets that have extreme low probability of being triadic and only focus on the triples where there is uncertainty whether they are triadic or not.

\begin{figure}[!ht]
	\centering
	\includegraphics[width=15cm,keepaspectratio]{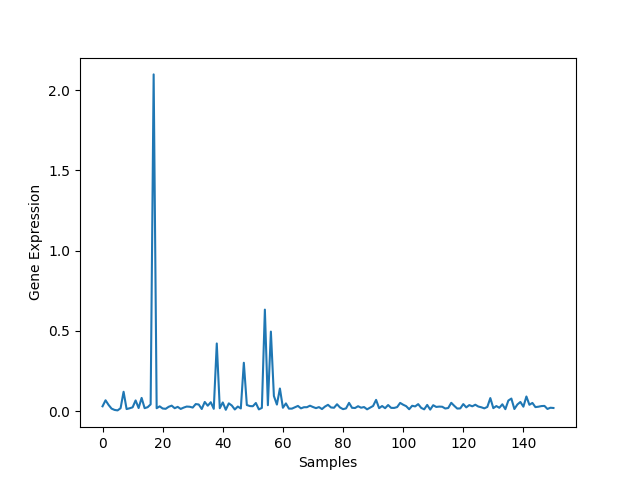}
	\caption{Samples of gene \textit{TGIF2LX}: Strength of the expression of the gene plotted against each individual patient tested. The biggest outlier is at position $17$ with a value of $2.097$, whereas the remaining values are smaller than $0.65$. }
 \label{S3}
\end{figure}

\begin{figure}
    \centering
    \includegraphics[width=0.5\linewidth]{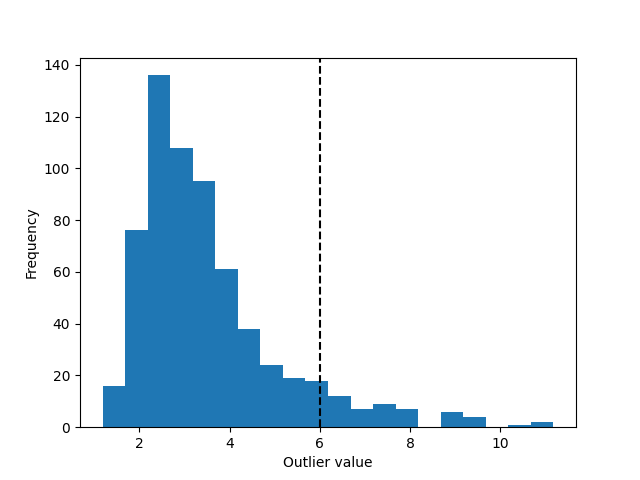}
    \caption{Cut-off for the top $50$ genes with high outlier values. The black vertical line indicates the cut-off. }
    \label{fig:cut_off}
\end{figure}

Our analysis of the data showed that several genes had an expression profile dominated by outliers that can strongly affect our analysis (see for an example the gene-expression profile of  gene \textit{TGIF2LX} in Figure \ref{S3}).
 For each gene $i$ we define an outlier score $O_i$  given by 
\bea
O_{i}=\max_{\alpha}\frac{|e_{i\alpha}-\Avg{e_i}|}{\sqrt{\Avg{e_i^2}-\Avg{e_i}^2}}
\label{O}
\eea
where $e_{i\alpha}$ is the gene-expression of gene $i$ in sample/patient $\alpha$, $\Avg{e_i^k}=\sum_{\alpha}e_{i\alpha}^k/Q$ where $Q$ is the number of samples.
The maximum z-score $O_i$ defined in Eq.(\ref{O}) is a good measure to detect outliers. We thus remove the top $50$ genes that score highest according to the $O$-score, corresponding to  a cut-off  $O\simeq 6$ as can be retrieved from a Figure \ref{fig:cut_off}.
These are the outliers for which triples containing them are removed: [\textit{TGIF2LX}, \textit{DBX2}, \textit{GSX1}, \textit{POU4F2}, \textit{EVX2}, \textit{EN2}, \textit{ONECUT1}, \textit{PAX2}, \textit{POU1F1},
\textit{MNX1}, \textit{HOXC10}, \textit{BARX2}, \textit{HOXD12}, \textit{HOXC12}, \textit{NKX2-5}, \textit{FOXC2}, \textit{HOXC11},
\textit{PROX1}, \textit{HMX2}, \textit{FOXF1}, \textit{HOXC8}, \textit{NR0B1}, \textit{PAX1}, \textit{GSC}, \textit{DMRT3}, \textit{HOXD13},
\textit{POU3F3}, \textit{FOXA2}, \textit{HOXD9}, \textit{HNF4G}, \textit{FOXF2}, \textit{SP8}, \textit{HMX3}, \textit{TLX1}, \textit{FOXD1},
\textit{FOXA1}, \textit{PHOX2B}, \textit{SOX21}, \textit{DMRT1}, \textit{FOXQ1}, \textit{VSX1}, \textit{HOXC6}, \textit{EGR4}, \textit{NKX2-1},
\textit{ZIC2}, \textit{LBX1}, \textit{RXRG}, \textit{EMX2}, \textit{SPZ1}
]. 
Having removed the outlier genes from our analysis, we test for triadic interactions using the  randomisation null model while the  Gaussian null model, being more sensitive to the outliers is only used to exclude triples that are not significant.  

For each edge, we select only the five triples with the highest $\Theta_\Sigma$-value  because these will be the most likely candidates for a triadic interaction. This is consistent with what we find for the synthetic data where triples containing the same edge can have higher $\Theta_\Sigma$ or $p_\Sigma$ values but only the highest ranking is the true triadic triple. To verify that these are triadic and the dependencies do not stem from edges connecting the regulatory node with the other two nodes in the triple, we repeat the analysis for a subset of these triples with the Gaussian version of the null model realisations. 
\subsection{Supplementary results}

\begin{figure}
    \centering
    \includegraphics[width=0.9\linewidth]{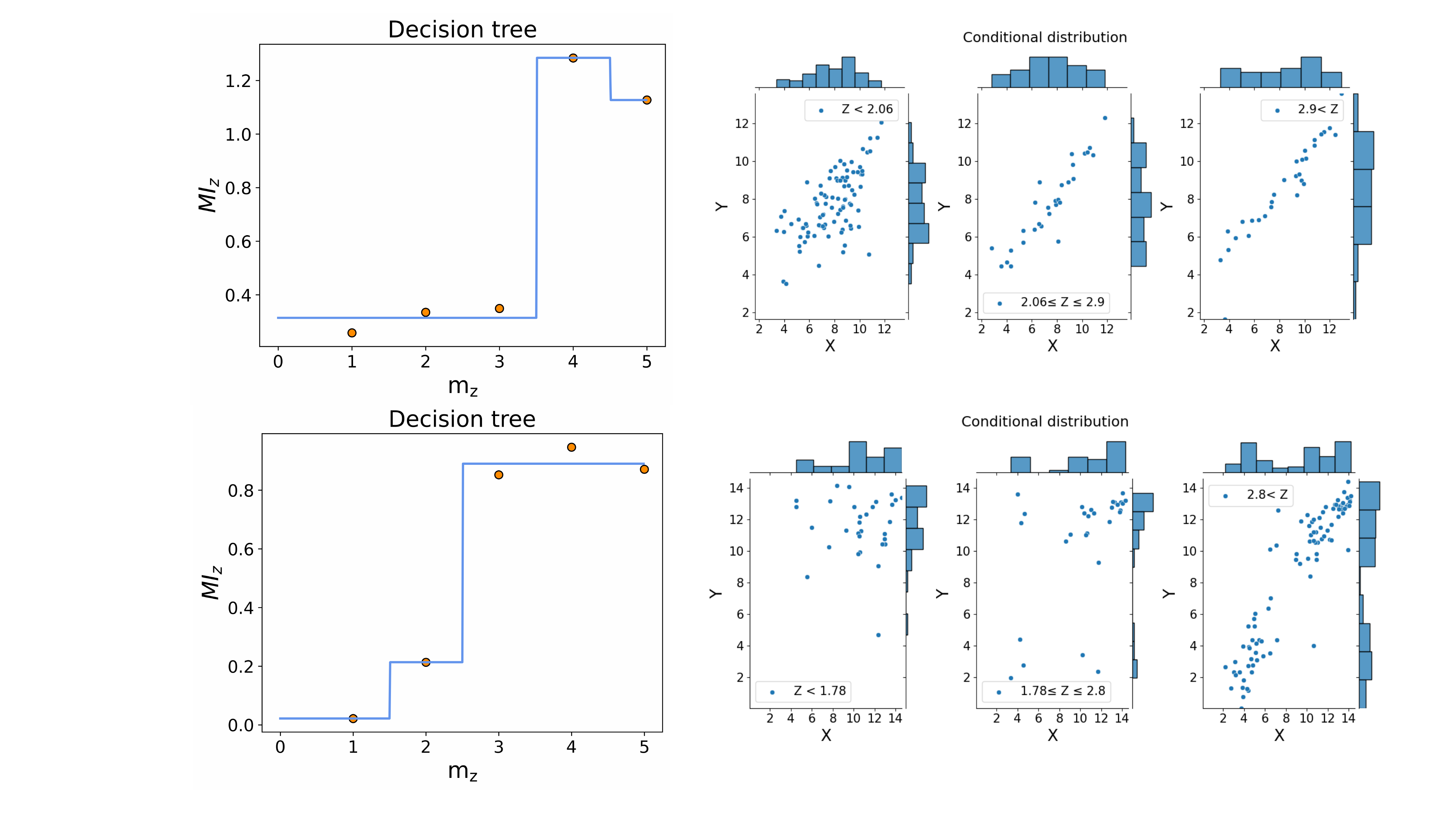}
    \caption{Decision trees belonging to the conditional distribution of the triples shown in Figure 8. Entropy for the top triple is $S=0.64$, entropy for the bottom triple is $S=0.60$.}
    \label{fig:7b}
\end{figure}

\begin{figure}
    \centering
    \includegraphics[width=0.5\linewidth]{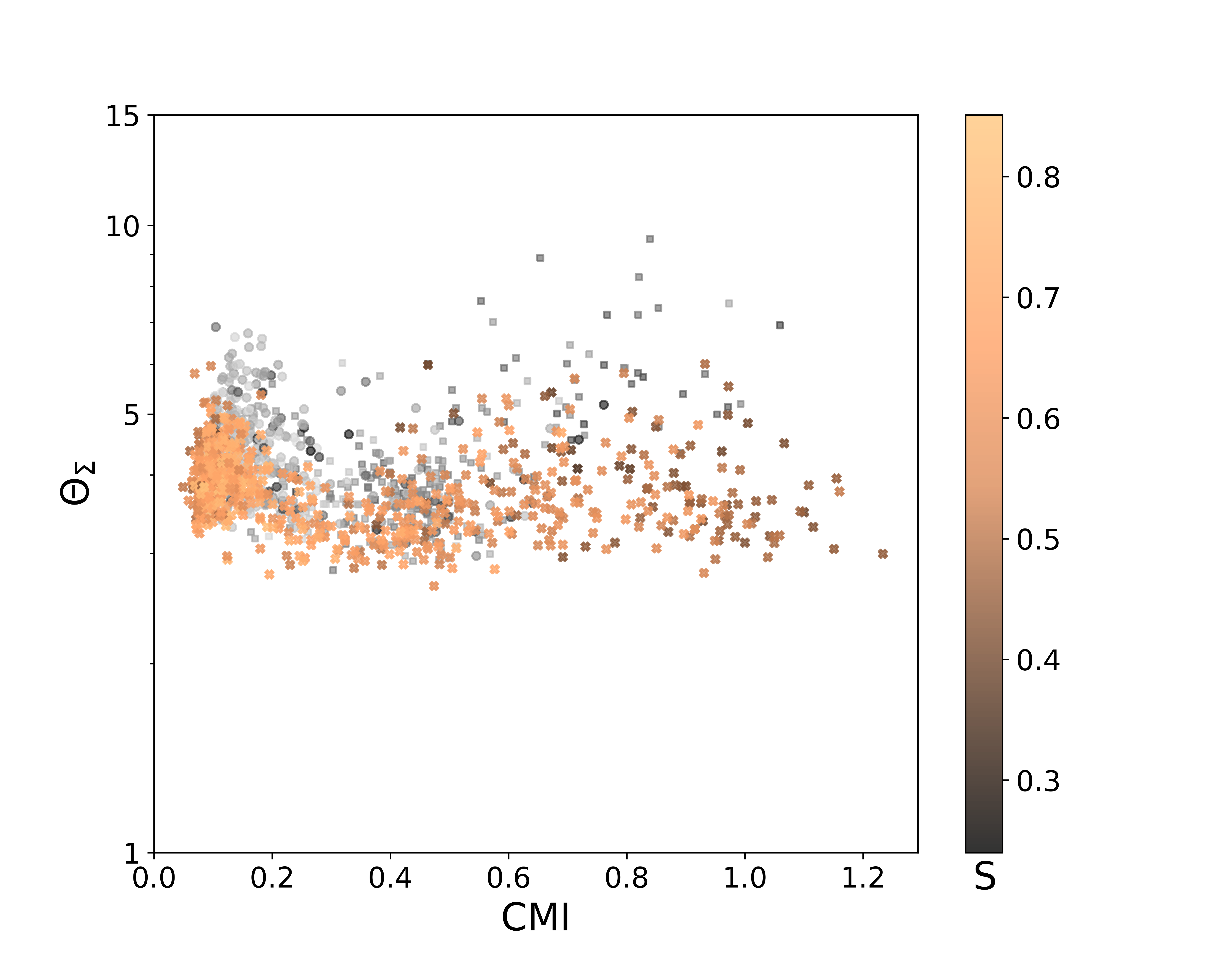}
    \caption{{The role of the Gaussian null model in screening out triples in the TRIM pipeline applied to the AML gene expression data.
   The scatter plot displays $\Theta_\Sigma$ (Y-axis) verse the CMI (X-axis). The colour corresponds to the value of our entropic score $S$. Here only triples that have $p$-value 0.001 or less in the randomization null model are shown.  
    Crosses (in colour) are triples that have $p$-value higher than 0.001 in the Gaussian null model and therefore they are disregarded from the analysis according to the TRIM pipeline.
   Symbols in gray include the triples that are retained and that figure in Figure 8 of the main body of the paper. Specifically these include triples whose links appear in the MST (circles), and  triples involving genes with biological relevance (squares).}}
    \label{fig:Gaussian}
\end{figure}

The analysis reported in Figure 8 of the main-body of the paper focuses on the results obtained by using the $\Sigma$ statistics.   
As supplementary material, in Figure $\ref{fig:7b}$  we report the decision trees corresponding to the conditional distribution of the triples shown in Figure 8 {{and  in Figure $\ref{fig:Gaussian}$ we report the triples screened out according to the TRIM pipeline because they have a $p$-value greater than $0.001$ according to the Gaussian null model.}}

In Table $\ref{tab1}$ we  use  Kendall's Tau to compare the ranking  obtained using the two  $\Theta_{\Sigma}$ and $\Theta_T$. The Kendall's Tau is higher than $0.7$ for both random and Gaussian null models so those measures are very similar. 

\begin{figure}[h!]
    \centering
    \includegraphics[width=\linewidth]{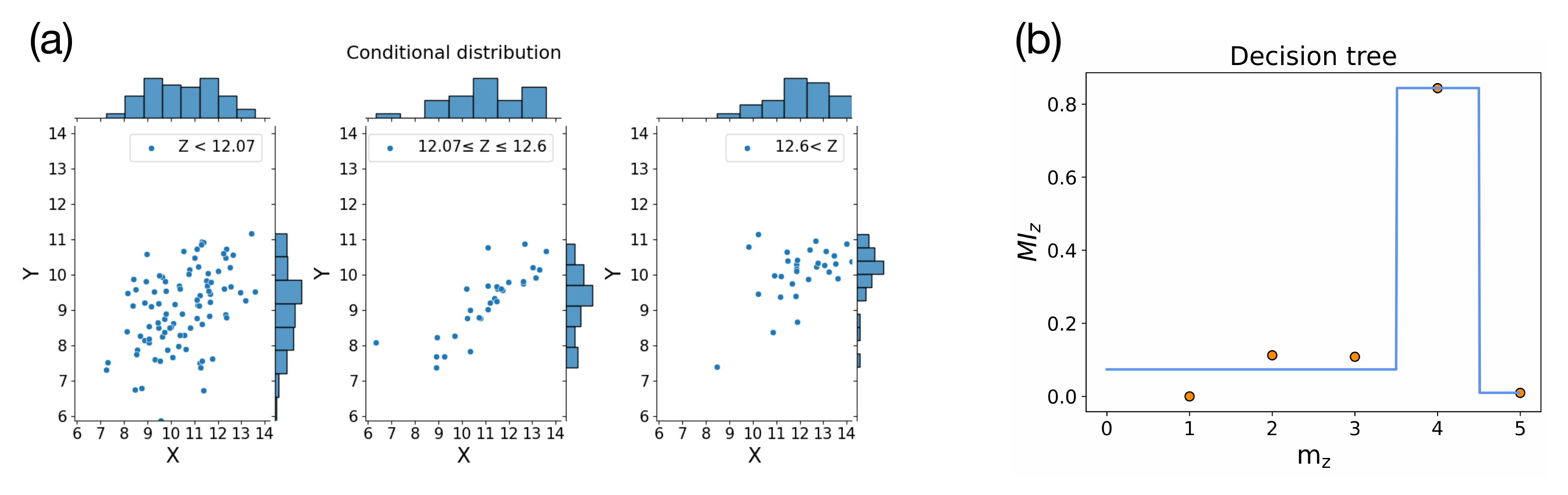}
    \caption{An example of non-monotonic triadic interaction. The figure (a) display the conditional joined distributions for the triple  X = \textit{BCL6}, Y = \textit{PPARD}, Z = \textit{IRF5}, $\Theta_\Sigma$ = 4.23, $p_\Sigma$ = 0.001, $\Sigma$ = 0.32, $S=0.71$,  and (b) the decision tree of the same triple.  }
    \label{fig:nonmon}
\end{figure}

\begin{table}[]
    \centering
    \caption{Comparison of the two measures $\Theta_\Sigma$ and $\Theta_T$  using Kendall-Tau analysis {for the Top 50 Triples in our Gene Expression Analysis.}}
    \begin{tabular}{ccc}
        \hline\hline
        &  statistic & $p$-value\\
        \hline
        Gaussian: $\Theta_\Sigma-\Theta_T$ & $0.864$ & $8.12 \times 10^{-19}$\\
        Random: $\Theta_\Sigma-\Theta_T$ & $0.703$ & $5.93 \times 10^{-13}$\\
        \hline\hline
    \end{tabular}
    \label{tab1}
\end{table}
As mentioned in the main text the TRIM algorithm can be potentially used to detect triadic interactions with non-monotonic behaviour.
The joint conditional distributions and the decision tree of one exemplary non-monotonic triples is shown in Figure $\ref{fig:nonmon}$. \\

The TRIM algorithm was also used to mine triadic interactions among all the triples containing genes that have been found to be highly relevant for triadic interactions from a biological point of view. These genes are \textit{HOXA1}, \textit{HOXA2}, \textit{HOXA3}, \textit{HOXA4}, \textit{HOXA5}, \textit{HOXA6}, \textit{HOXA7}, \textit{HOXA8}, \textit{HOXA9}, 
\textit{HOXB1}, \textit{HOXB2}, \textit{HOXB3}, \textit{HOXB4}, \textit{HOXB5}, \textit{HOXB6}, \textit{HOXB7}, \textit{HOXB8}, \textit{HOXB9}, 
\textit{PBX3}, \textit{MEIS1} \cite{Alharbi2013}, \cite{Guo2017}, \cite{Xiang2022}. There are $125,244$ triples of this type, 
260 have been found to be candidates after applying the TRIM pipeline. Especially we observe that the method is able to give a high significance score to the  triple \textit{HOXA9}, \textit{HOXA3}, \textit{MEIS1} that has $p_\Sigma$-value $0.001$ and $\Theta_\Sigma$ greater than 6 which is a triple of  overexpressed genes in AML \cite{Alharbi2013}.  
This triple together with the highest ranking biological triples by $\Theta_\Sigma$ are shown in Table $\ref{tab:bio}$.

Moreover in Table $\ref{tab:top}$ we report the data for the top $50$ candidates of triadic interactions obtained using our procedure.

Finally in Table $\ref{tab:lit}$ we provide the list of the literature on the genes involved on the top scoring triples detected by our analysis of the AML gene-expression data.

\begin{table*}[]
    \centering
    \caption{Highest ranking Biological Triples by $\Theta_\Sigma$}
    \label{tab:bio}
    \begin{ruledtabular}
    \begin{tabular}{ccccccccccc}
        Node 1 & Node 2 & Regulatory Node & CMI & $\Theta_\Sigma$ & $\Theta_T$ & $\Sigma$ & $P_\Sigma$ & $P_T$ & $S$\\\hline
         \textit{HOXA3}  & \textit{HOXA5} & \textit{PBX3}   & 0.820316 & 8.265877 & 8.048662 & 0.437625 & 0.001 & 0.001 & 0.511452\\
         \textit{HOXA5}  & \textit{HOXA3} & \textit{LMX1B}  & 0.973260 & 7.506890 & 6.763280 & 0.401504 & 0.001 & 0.001 & 0.631465\\
         \textit{HOXB5}  & \textit{HOXB6} & \textit{MEIS1}  & 0.819451 & 7.205339 & 6.846796 & 0.427434 & 0.001 & 0.001 & 0.501928\\
         \textit{HOXA2}  & \textit{HOXA5} & \textit{MEIS1}  & 0.573735 & 7.019063 & 5.996416 & 0.367280 & 0.001 & 0.001 & 0.596266\\
         \textit{HOXA2}  & \textit{HOXA5} & \textit{MEF2A}  & 0.704308 & 6.448953 & 6.374430 & 0.357321 & 0.001 & 0.001 & 0.618474\\
         \textit{HOXA3}  & \textit{HOXA2} & \textit{MEF2A}  & 0.736876 & 6.229782 & 5.449352 & 0.357688 & 0.001 & 0.001 & 0.618665\\
         \textit{HOXA9}  & \textit{HOXA3} & \textit{MEIS1}  & 0.612700 & 6.147447 & 5.840250 & 0.376088 & 0.001 & 0.001 & 0.516719\\
         \textit{HOXA1}  & \textit{HOXB7} & \textit{DLX3}   & 0.319020 & 6.031555 & 5.996593 & 0.377072 & 0.001 & 0.001 & 0.719936\\
         \textit{HOXA3}  & \textit{HOXA7} & \textit{LMX1B}  & 0.796522 & 5.930516 & 5.361882 & 0.315926 & 0.001 & 0.001 & 0.575184\\
         \textit{HOXA9}  & \textit{HOXA7} & \textit{LMX1B}  & 0.795180 & 5.921237 & 6.062873 & 0.351938 & 0.001 & 0.001 & 0.520730\\
         \textit{HOXB3}  & \textit{HOXB6} & \textit{HOXB7}  & 0.382231 & 5.755689 & 4.911239 & 0.320005 & 0.001 & 0.001 & 0.636879\\
         \textit{HOXB6}  & \textit{HOXB3} & \textit{MAX}    & 0.711938 & 5.676961 & 6.039347 & 0.309040 & 0.001 & 0.001 & 0.535384\\
         \textit{HOXA3}  & \textit{HOXA2} & \textit{MEIS1}  & 0.632230 & 5.643140 & 5.226696 & 0.343473 & 0.001 & 0.001 & 0.650853\\	
         \textit{HOXB1}  & \textit{PBX3}  & \textit{DMRTC2} & 0.200534 & 5.582683 & 5.388206 & 0.305488 & 0.001 & 0.001 & 0.566194\\
         \textit{HOXB5}  & \textit{HOXB3} & \textit{HOXA7}  & 0.504192 & 5.464608 & 5.008018 & 0.320871 & 0.001 & 0.001 & 0.519581\\	
         \textit{HOXB5}  & \textit{HOXB6} & \textit{HOXA3}  & 0.719784 & 5.333075 & 5.807730 & 0.357958 & 0.001 & 0.001 &0.5175458\\	
         \textit{HOXA2}  & \textit{HOXA5} & \textit{LMX1B}  & 0.685202 & 5.254939 & 5.041743 & 0.314368 & 0.001 & 0.001 & 0.710364\\	
         \textit{HOXA3}  & \textit{HOXA2} & \textit{HOXB3}  & 0.614658 & 5.212560 & 5.482464 & 0.325482 & 0.001 & 0.001 & 0.669220\\
         \textit{HOXB2}  & \textit{HOXB3} & \textit{SPI1}   & 0.992869 & 5.194840 & 5.064013 & 0.338784 & 0.001 & 0.001 & 0.542340\\
         \textit{HOXA1}  & \textit{HOXB8} & \textit{RUNX3}  & 0.176414 & 5.148190 & 4.848085 & 0.286748 & 0.001 & 0.001 & 0.626724\\
         \textit{HOXA5}  & \textit{HOXA2} & \textit{SIX3}   & 0.697338 & 5.129497 & 5.577772 & 0.309603 & 0.001 & 0.001 & 0.542347\\
         \textit{HOXB7}  & \textit{HOXB6} & \textit{MLXIPL} & 0.511277 & 5.113759 & 5.390683 & 0.361182 & 0.001 & 0.001 & 0.542626\\
         \textit{HOXA7}  & \textit{PBX3}  & \textit{POU4F1} & 0.555084 & 5.111917 & 5.837126 & 0.312872 & 0.001 & 0.001 & 0.586764\\
         \textit{HOXB8}  & \textit{HOXA1} & \textit{DLX3}   & 0.206524 & 5.058267 & 5.102255 & 0.278793 & 0.001 & 0.001 & 0.609568\\
         \textit{HOXB6}  & \textit{HOXB7} & \textit{STAT5B} & 0.563708 & 5.048295 & 4.784504 & 0.355652 & 0.001 & 0.001 & 0.569296\\
         \textit{HOXA3}  & \textit{HOXA2} & \textit{HOXB8}  & 0.704156 & 4.982283 & 4.596447 & 0.326155 & 0.001 & 0.001 & 0.5670250\\
         \textit{HOXB1}  & \textit{PBX3}  & \textit{SREBF1} & 0.238224 & 4.964400 & 3.933486 & 0.290385 & 0.001 & 0.001 & 0.5993946\\
         \textit{MEIS1}  & \textit{HOXA3} & \textit{NR6A1}  & 0.484028 & 4.787769 & 4.564621 & 0.384765 & 0.001 & 0.001 & 0.5612223\\
         \textit{HOXA3}  & \textit{HOXA2} & \textit{ERG}    & 0.854079 & 4.775633 & 4.719983 & 0.307792 & 0.001 & 0.001 & 0.5308264\\
         \textit{HOXB3}  & \textit{HOXB5} & \textit{IRX1}   & 0.681236 & 4.762596 & 4.095362 & 0.287235 & 0.001 & 0.001 & 0.5965832\\
         \textit{HOXB2}  & \textit{HOXB6} & \textit{HOXB7}  & 0.349210 & 4.660569 & 3.497816 & 0.316489 & 0.001 & 0.002 & 0.6341667\\
         \textit{HOXB5}  & \textit{HOXB6} & \textit{HOXA9}  & 0.728788 & 4.625061 & 3.950498 & 0.325499 & 0.001 & 0.001 & 0.5515168\\
         \textit{HOXA9}  & \textit{HOXB7} & \textit{MLXIPL} & 0.168949 & 4.618047 & 3.982716 & 0.289461 & 0.001 & 0.001 & 0.6315528\\
         \textit{HOXB3}  & \textit{HOXB5} & \textit{HOXA5}  & 0.547398 & 4.574438 & 5.077232 & 0.286324 & 0.001 & 0.001 & 0.6519540\\
         \textit{HOXB1}  & \textit{HOXA5} & \textit{MAFA}   & 0.258676 & 4.553432 & 4.648329 & 0.279503 & 0.001 & 0.001 & 0.6090695\\
         \textit{HOXA1}  & \textit{HOXB6} & \textit{ATF4}   & 0.371474 & 4.544329 & 4.315940 & 0.299968 & 0.001 & 0.001 & 0.6979978\\
         \textit{MEIS1}  & \textit{HOXA5} & \textit{HEYL}   & 0.488968 & 4.526637 & 3.193835 & 0.382446 & 0.001 & 0.002 & 0.6221348\\
         \textit{HOXB2}  & \textit{HOXB5} & \textit{IRX1}   & 0.661687 & 4.473764 & 4.903444 & 0.295175 & 0.001 & 0.001 & 0.6057017\\
         \textit{HOXB7}  & \textit{HOXB6} & \textit{HOXA9}  & 0.346855 & 4.444519 & 3.487488 & 0.328127 & 0.001 & 0.002 & 0.5408434\\
         \textit{HOXB3}  & \textit{HOXB7} & \textit{CEBPA}  & 0.456098 & 4.437850 & 5.265165 & 0.302595 & 0.001 & 0.001 & 0.7035531\\
         \textit{HOXB8}  & \textit{HOXA1} & \textit{PLAGL1} & 0.174812 & 4.435822 & 4.227541 & 0.259337 & 0.001 & 0.001 & 0.6397053\\
         \textit{PBX3}   & \textit{HOXA3} & \textit{DLX4}   & 0.376259 & 4.341928 & 3.383559 & 0.294993 & 0.001 & 0.003 & 0.5599292\\
         \textit{HOXA1}  & \textit{HOXA5} & \textit{RARA}   & 0.564074 & 4.283333 & 4.080688 & 0.347871 & 0.001 & 0.002 & 0.7039146\\
         \textit{HOXB7}  & \textit{HOXB5} & \textit{CEBPA}  & 0.523503 & 4.248256 & 4.572174 & 0.316662 & 0.001 & 0.001 & 0.5829643\\
         \textit{HOXB8}  & \textit{HOXA9} & \textit{IRX5}   & 0.143996 & 4.244078 & 3.743148 & 0.255930 & 0.001 & 0.002 & 0.5426364\\
         \textit{HOXA7}  & \textit{PBX3}  & \textit{MEIS2}  & 0.481629 & 4.229047 & 4.704656 & 0.291258 & 0.001 & 0.001 & 0.5720851\\
         \textit{HOXA1}  & \textit{HOXA3} & \textit{JDP2}   & 0.462743 & 4.227947 & 4.378695 & 0.349784 & 0.001 & 0.001 & 0.6529631\\
         \textit{MEIS1}  & \textit{PBX3}  & \textit{FOXI1}  & 0.399453 & 4.217408 & 3.765785 & 0.351739 & 0.001 & 0.001 & 0.5882254\\
         \textit{HOXB7}  & \textit{HOXB6} & \textit{HOXA2}  & 0.396262 & 4.213586 & 3.643030 & 0.323173 & 0.001 & 0.001 & 0.6443882\\
         \textit{HOXA2}  & \textit{HOXB7} & \textit{PBX1}   & 0.297987 & 4.206895 & 2.752253 & 0.303415 & 0.001 & 0.005 & 0.6995975\\
    \end{tabular}
    \end{ruledtabular}
\end{table*}
\begin{table*}[]
    \centering
    \caption{Triples with the $50$ highest $\Theta_\Sigma$-values. Chosen from the analysed triples where only the five triples with highest $\Theta_\Sigma$-values for each edge are considered and after the removal of $50$ outlier genes.}
    \label{tab:top}
    \setlength{\tabcolsep}{4pt} 
    \begin{ruledtabular}
    \begin{tabular}{cccccccccc}
        Node 1 & Node 2 & Reg. Node & CMI & $\Theta_\Sigma$ & $\Theta_T$  & $\Sigma$ & $P_\Sigma$ & $P_T$ & $S$\\
        \hline
        \textit{GFI1B}  & \textit{HMG20B}  & \textit{MAFG}      & 0.159214 & 6.728150 & 6.919335 & 0.239888 & 0.001 & 0.001 & 0.656489\\
        \textit{ESR1}   & \textit{ESRRA}   & \textit{CUX1}      & 0.136866 & 6.633394 & 6.387613 & 0.257335 & 0.001 & 0.001 & 0.740265\\
        \textit{ATF2}   & \textit{CREB3L2} & \textit{NKX2-3}    & 0.183120 & 6.597125 & 6.372863 & 0.292662 & 0.001 & 0.001 & 0.699403\\
        \textit{AR}     & \textit{ERG}     & \textit{DLX3}      & 0.181392 & 6.419430 & 6.294301 & 0.264732 & 0.001 & 0.001 & 0.670266\\
        \textit{GMEB1}  & \textit{GMEB2}   & \textit{GABPA}     & 0.160774 & 6.393412 & 6.419077 & 0.249189 & 0.001 & 0.001 & 0.648156\\
        \textit{TP53}   & \textit{TOPORS}  & \textit{RFX1}      & 0.132448 & 6.245761 & 6.067981 & 0.245629 & 0.001 & 0.001 & 0.636249\\
        \textit{SIX3}   & \textit{NR4A3}   & \textit{PPARD}     & 0.126805 & 6.157518 & 5.990863 & 0.226794 & 0.001 & 0.001 & 0.626979\\
        \textit{CREM}   & \textit{CREB3}   & \textit{GABPA}     & 0.145051 & 6.013678 & 6.087598 & 0.217777 & 0.001 & 0.001 & 0.705945\\
        \textit{NR1H2}  & \textit{RXRB}    & \textit{FOXO1}     & 0.210384 & 5.997208 & 5.846367 & 0.283394 & 0.001 & 0.001 & 0.622814\\
        \textit{AIRE}   & \textit{GMEB1}   & \textit{ZSCAN16}   & 0.139642 & 5.986620 & 5.997924 & 0.230049 & 0.001 & 0.001 & 0.742193\\
        \textit{MAFG}   & \textit{NFE2}    & \textit{SMARCC1}   & 0.133617 & 5.963117 & 5.929599 & 0.233347 & 0.001 & 0.001 & 0.675628\\
        \textit{MAFG}   & \textit{NFE2}    & \textit{MBD2}      & 0.129252 & 5.962294 & 5.873812 & 0.233020 & 0.001 & 0.001 & 0.656413\\
        \textit{EGR2}   & \textit{HOXB2}   & \textit{HOXA9}     & 0.156973 & 5.865105 & 5.867066 & 0.228347 & 0.001 & 0.001 & 0.688156\\
        \textit{MEIS1}  & \textit{PBX1}    & \textit{ZEB1}      & 0.188129 & 5.842200 & 5.931817 & 0.265427 & 0.001 & 0.001 & 0.611296\\
        \textit{WT1}    & \textit{SOX12}   & \textit{ARX}       & 0.172959 & 5.825481 & 5.423547 & 0.310742 & 0.001 & 0.001 & 0.622995\\
        \textit{MYOD1}  & \textit{TCF12}   & \textit{RFX1}      & 0.134786 & 5.809604 & 5.792604 & 0.202536 & 0.001 & 0.001 & 0.574806\\
        \textit{KDM2B}  & \textit{ZNF784}  & \textit{TEAD1}     & 0.184460 & 5.808177 & 6.087558 & 0.237961 & 0.001 & 0.001 & 0.699574\\
        \textit{ELF1}   & \textit{TFDP1}   & \textit{KDM2B}     & 0.180421 & 5.783468 & 6.248786 & 0.216104 & 0.001 & 0.001 & 0.665745\\
        \textit{ZNF423} & \textit{EBF1}    & \textit{ZIC4}      & 0.134866 & 5.766551 & 5.550101 & 0.245472 & 0.001 & 0.001 & 0.714254\\
        \textit{TP53}   & \textit{IRF5}    & \textit{BCL6}      & 0.185771 & 5.751306 & 6.407961 & 0.206465 & 0.001 & 0.001 & 0.612062\\
        \textit{ESR1}   & \textit{XBP1}    & \textit{HEY2}      & 0.217111 & 5.741680 & 5.731264 & 0.262501 & 0.001 & 0.001 & 0.707043\\
        \textit{MEIS3}  & \textit{PBX3}    & \textit{DRGX}      & 0.116614 & 5.726887 & 5.518357 & 0.225603 & 0.001 & 0.001 & 0.640716\\
        \textit{FOS}    & \textit{ELK1}    & \textit{RORA}      & 0.128186 & 5.714594 & 5.384268 & 0.256372 & 0.001 & 0.001 & 0.698619\\
        \textit{LMX1B}  & \textit{GBX1}    & \textit{MSX2}      & 0.114587 & 5.677847 & 5.501699 & 0.227926 & 0.001 & 0.001 & 0.626511\\
        \textit{CEBPA}  & \textit{CEBPB}   & \textit{SMARCC1}   & 0.149690 & 5.675830 & 5.864768 & 0.232468 & 0.001 & 0.001 & 0.629392\\
        \textit{AR}     & \textit{HOXB13}  & \textit{IRX5}      & 0.127068 & 5.628973 & 5.607486 & 0.227873 & 0.001 & 0.001 & 0.677219\\
        \textit{LEF1}   & \textit{CDX1}    & \textit{CEBPG}     & 0.173025 & 5.582373 & 5.740356 & 0.204470 & 0.001 & 0.001 & 0.589337\\
        \textit{JUN}    & \textit{ATF2}    & \textit{IRF6}      & 0.160900 & 5.540512 & 5.706251 & 0.217531 & 0.001 & 0.001 & 0.743458\\
        \textit{LMX1B}  & \textit{FEV}     & \textit{CEBPE}     & 0.132053 & 5.459481 & 5.381255 & 0.217621 & 0.001 & 0.001 & 0.531703\\
        \textit{DLX6}   & \textit{HAND2}   & \textit{GFI1}      & 0.316684 & 5.447074 & 5.452548 & 0.409997 & 0.001 & 0.001 & 0.5666130\\
        \textit{EGR3}   & \textit{NFATC2}  & \textit{FOSB}      & 0.139154 & 5.444294 & 5.407629 & 0.230926 & 0.001 & 0.001 & 0.7545026\\
        \textit{ARNT2}  & \textit{EPAS1}   & \textit{STAT5B}    & 0.184543 & 5.421202 & 5.515499 & 0.269106 & 0.001 & 0.001 & 0.6476382\\
        \textit{HOMEZ}  & \textit{HMBOX1}  & \textit{POU3F2}    & 0.116605 & 5.417860 & 5.441199 & 0.199304 & 0.001 & 0.001 & 0.6313812\\
        \textit{AR}     & \textit{RREB1}   & \textit{LMO2}      & 0.114581 & 5.415712 & 5.247730 & 0.221936 & 0.001 & 0.001 & 0.7792968\\
        \textit{NPAS2}  & \textit{RORA}    & \textit{TOPORS}    & 0.139477 & 5.395178 & 5.542092 & 0.214479 & 0.001 & 0.001 & 0.7393471\\
        \textit{FOXP2}  & \textit{SP4}     & \textit{TBX3}      & 0.139812 & 5.377868 & 5.488782 & 0.217301 & 0.001 & 0.001 & 0.5843965\\
        \textit{GATA1}  & \textit{IKZF1}   & \textit{DPRX}      & 0.130265 & 5.357680 & 5.120758 & 0.233326 & 0.001 & 0.001 & 0.7042830\\
        \textit{NANOG}  & \textit{TET1}    & \textit{PURA}      & 0.142164 & 5.348978 & 5.556867 & 0.212450 & 0.001 & 0.001 & 0.7522146\\
        \textit{KDM2B}  & \textit{ZNF740}  & \textit{NOBOX}     & 0.103249 & 5.343459 & 5.074403 & 0.205580 & 0.001 & 0.001 & 0.7296110\\
        \textit{PITX2}  & \textit{LHX3}    & \textit{TBX5}      & 0.156647 & 5.315290 & 4.504254 & 0.193797 & 0.001 & 0.001 & 0.6662439\\
        \textit{AR}     & \textit{ERG}     & \textit{DDIT3}     & 0.138062 & 5.307119 & 5.147755 & 0.236228 & 0.001 & 0.001 & 0.6881829\\
        \textit{KDM2B}  & \textit{ZNF232}  & \textit{MZF1}      & 0.102529 & 5.306776 & 5.161613 & 0.197083 & 0.001 & 0.001 & 0.6727312\\
        \textit{ARID2}  & \textit{SMARCC2} & \textit{VDR}       & 0.106636 & 5.273111 & 5.122535 & 0.203007 & 0.001 & 0.001 & 0.6993566\\
        \textit{ESR1}   & \textit{NR2C2}   & \textit{HOXD8}     & 0.129780 & 5.247737 & 5.008668 & 0.234352 & 0.001 & 0.001 & 0.7039140\\
        \textit{LMO2}   & \textit{TAL1}    & \textit{ZNF32}     & 0.105264 & 5.224616 & 5.238358 & 0.182179 & 0.001 & 0.001 & 0.7148965\\
        \textit{ARNT}   & \textit{EPAS1}   & \textit{REST}      & 0.121260 & 5.167161 & 5.191644 & 0.214579 & 0.001 & 0.001 & 0.7482859\\
        \textit{LMO2}   & \textit{NHLH1}   & \textit{TBX19}     & 0.102305 & 5.166281 & 5.103774 & 0.196432 & 0.001 & 0.001 & 0.7178833\\
        \textit{E2F3}   & \textit{TFDP1}   & \textit{BPTF}      & 0.184846 & 5.118170 & 4.863077 & 0.203638 & 0.001 & 0.001 & 0.6571473\\
        \textit{SMAD4}  & \textit{SMAD5}   & \textit{HOXB6}     & 0.443231 & 5.113130 & 5.490316 & 0.374981 & 0.001 & 0.001 & 0.6213580\\
        \textit{MITF}   & \textit{PAX3}    & \textit{MLXIPL}    & 0.131645 & 5.102654 & 4.907555 & 0.173291 & 0.001 & 0.001 & 0.6373616\\
    \end{tabular}
    \end{ruledtabular}
\end{table*}

\begin{table*}
    \centering
    \caption{Genes included in the $50$ triples with highest $\Theta_\Sigma$-value. The list includes genes relevant to AML found in literature (left) and a list of triples (right). Genes with relevance to AML have been bolded.}
    \label{tab:lit}
    \begin{ruledtabular}
        \begin{tabular}{clccc}
            Gene & Reference & Gene 1 &  Gene 2 & Regulatory Gene \\
            \hline
            \textit{ARID2} & \cite{Bluemn2022} T. Bluemn, et al. (2022) & \textbf{\textit{GFI1B}} & \textit{HMG20B} & \textit{MAFG}\\
            \textit{ARNT} & \cite{Williams2010} S. Williams, et al. (2010) & \textbf{\textit{ESR1}} & \textit{ESRRA} & \textbf{\textit{CUX1}} \\
            \textit{BCL6} & \cite{Kawabata2021} K. Kawabata, et al. (2021) & \textit{ATF2} & \textit{CREB3L2} & \textbf{\textit{NKX2-3}}\\
            \textit{BPTF} & \cite{Radzisheuskaya2023} A. Radzisheuskaya, et al. (2023) & \textit{AR} & \textbf{\textit{ERG}} & \textit{DLX3}\\
            \textit{CEBPA} & \cite{Fasan2014} A. Fasan, et al. (2014) & \textit{GMEB1} & \textit{GMEB2} & \textit{GABPA}\\
            \textit{CEBPB} & \cite{Fu2022} W. Fu, et al. (2022) & \textbf{\textit{TP53}} & \textit{TOPORS} & \textit{RFX1}\\
            \textit{CEBPE} & \cite{Kening2019} Li, K. et al. (2019) & \textit{SIX3} & \textbf{\textit{NR4A3}} & \textit{PPARD}\\
            \textit{CEBPG} & \cite{Jiang2021} Y. Jiang, et al. (2021) & \textit{CREM} & \textbf{\textit{CREB3}} & \textit{GABPA}\\
            \textit{CREB3} & \cite{Feng2020} S. Feng, et al. (2020) & \textit{NR1H2} & \textit{RXRB} & \textbf{\textit{FOXO1}}\\
            \textit{CUX1} & \cite{McNerney2013} M. E. McNerney, et al. (2013) & \textit{AIRE} & \textit{GMEB1} & \textit{ZSCAN16}\\
            \textit{ELF1} & \cite{Varghese2025} P. Varghese, et al. (2025) & \textit{MAFG} & \textbf{\textit{NFE2}} & \textit{SMARCC1}\\
            \textit{ELK1} & \cite{Guo2023} D. Guo, et al. (2023) & \textit{MAFG} & \textbf{\textit{NFE2}} & \textbf{\textit{MBD2}}\\
            \textit{EPAS1} & \cite{Wang2023} S. Wang, et al. (2023) & \textit{EGR2} & \textbf{\textit{HOXB2}} & \textbf{\textit{HOXA9}}\\
            \textit{ERG} & \cite{Marcucci2005} G. Marcucci, et al. (2005) & \textbf{\textit{MEIS1}} & \textit{PBX1} & \textbf{\textit{ZEB1}}\\
            \textit{ESR1} & \cite{Roma2020} A. Roma, et al. (2020) & \textbf{\textit{WT1}} & \textbf{\textit{SOX12}} & \textit{ARX}\\
            \textit{FEV} & \cite{Zhang2022} J. Zhang, et al. (2022) & \textbf{\textit{MYOD1}} & \textit{TCF12} & \textit{RFX1}\\
            \textit{FOS} & \cite{Yang2024} F. Yang, et al. (2024) & \textbf{\textit{KDM2B}} & \textit{ZNF784} & \textit{TEAD1}\\
            \textit{FOSB} & \cite{Luan2022} SH. Luan, et al. (2022) & \textbf{\textit{ELF1}} & \textit{TFDP1} & \textbf{\textit{KDM2B}}\\
            \textit{FOXO1} & \cite{Lin2014} S. Lin, et al. (2014) & \textit{ZNF423} & \textit{EBF1} & \textit{ZIC4}\\
            \textit{GATA1} & \cite{Ayala2009} R. Ayala, et al. (2009) & \textbf{\textit{TP53}} & \textit{IRF5} & \textbf{\textit{BCL6}}\\
            \textit{GFI1} & \cite{Moeroey2015} T. M{\"o}r{\"o}y, et al. (2015) & \textbf{\textit{ESR1}} & \textit{XBP1} & \textit{HEY2}\\
            \textit{HOXA9} &  \cite{aryal2023} S. Aryal, et al. (2023) & \textit{MEIS3} & \textbf{\textit{PBX3}} & \textit{DRGX}\\
             \textit{HOXB13} & \cite{chu2018} Y. Chu, et al. (2018) & \textbf{\textit{FOS}} & \textbf{\textit{ELK1}} & \textbf{\textit{RORA}}\\
             \textit{HOXB2} & \cite{LINDBLAD2015} O. Lindblad, et al. (2015) & \textit{LMX1B} & \textit{GBX1} & \textit{MSX2}\\
             \textit{HOXB6} & \cite{Giampaolo2002vw}  A. Giampaolo, et al. (2002) & \textbf{\textit{CEBPA}} & \textbf{\textit{CEBPB}} & \textit{SMARCC1}\\
             \textit{IKZF1} & \cite{Eckardt2023} J. Eckardt, et al. (2023) & \textit{AR} & \textbf{\textit{HOXB13}} & \textbf{\textit{IRX5}}\\
             \textit{IRX5} &  \cite{ijms23063192} S. Nagel, et al. (2022) & \textbf{\textit{LEF1}} & \textit{CDX1} & \textbf{\textit{CEBPG}}\\
             \textit{JUN} & \cite{Zhou2017} C. Zhou, et al. (2017) & \textbf{\textit{JUN}} & \textit{ATF2} & \textit{IRF6}\\
             \textit{KDM2B} & \cite{Boom2016} V. van den Boom, et al. (2016) & \textit{LMX1B} & \textit{FEV} & \textbf{\textit{CEBPE}}\\
             \textit{LEF1} & \cite{Feder2020} K. Feder, et al. (2020) & \textit{DLX6} & \textit{HAND2} & \textbf{\textit{GFI1}}\\
             \textit{LMO2} & \cite{Lu2023} L. Lu, et al. 2023 & \textit{EGR3} & \textbf{\textit{NFATC2}} & \textbf{\textit{FOSB}}\\
             \textit{MBD2} & \cite{Zhou2021} K. Zhou, et al. (2021) & \textit{ARNT2} & \textbf{\textit{EPAS1}} & \textbf{\textit{STAT5B}}\\
             \textit{MEIS1} & \cite{Thorsteinsdottir2001} U. Thorsteinsdottir, et al. (2001) & \textit{HOMEZ} & \textit{HMBOX1} & \textit{POU3F2}\\
             \textit{MYOD1} & \cite{Toyota2001} M. Toyota, et al. (2001) & \textit{AR} & \textit{RREB1} & \textbf{\textit{LMO2}}\\
             \textit{NFATC2} & \cite{Patterson2021} S. D. Patterson, et al. (2021) & \textbf{\textit{NPAS2}} & \textbf{\textit{RORA}} & \textit{TOPORS}\\
             \textit{NFE2} & \cite{Jutzi2019} J. S. Jutzi, et al. (2019) & \textit{FOXP2} & \textit{SP4} & \textit{TBX3}\\
             \textit{NKX2-3} & \cite{Nagel2021} S. Nagel, et al. (2021) & \textbf{\textit{GATA1}} & \textbf{\textit{IKZF1}} & \textit{DPRX}\\
             \textit{NPAS2} & \cite{Song2018} B. Song, et al. (2018) & \textit{NANOG} & \textbf{\textit{TET1}} & \textbf{\textit{PURA}}\\
             \textit{NR4A3} & \cite{Shih-Chiang2022} SC. Lin, et al. (2022) & \textbf{\textit{KDM2B}} & \textit{ZNF740} & \textit{NOBOX}\\
             \textit{PBX3} & \cite{Dickson2013} G. J. Dickson, et al. (2013) & \textit{PITX2} & \textit{LHX3} & \textit{TBX5}\\
             \textit{PURA} & \cite{Lezon-Geyda2001} K. Lezon-Geyda, et al. (2001) & \textit{AR} & \textbf{\textit{ERG}} & \textit{DDIT3}\\
             \textit{RORA} & \cite{Snider2019} C. Snider, et al. (2019) & \textbf{\textit{KDM2B}} & \textit{ZNF232} & \textit{MZF1}\\
             \textit{SMAD4} & \cite{Imai2001} Y. Imai, et al. (2001) & \textbf{\textit{ARID2}} & \textit{SMARCC2} & \textit{VDR}\\
             \textit{SOX12} & \cite{Wan2017} H. Wan, et al. (2017) & \textbf{\textit{ESR1}} & \textit{NR2C2} & \textit{HOXD8}\\
             \textit{STAT5B} & \cite{Maurer2019} B. Maurer, et al. (2019) & \textbf{\textit{LMO2}} & \textbf{\textit{TAL1}} & \textit{ZNF32}\\
             \textit{TAL1} & \cite{Wang2023} Z. Wang, et al. (2023) & \textbf{\textit{ARNT}} & \textbf{\textit{EPAS1}} & \textit{REST}\\
             \textit{TET1} & \cite{Wang2018} J. Wang, et al. (2018) & \textbf{\textit{LMO2}} & \textit{NHLH1} & \textit{TBX19}\\
             \textit{TP53} & \cite{Barbosa2019} K. Barbosa, et al. (2019) & \textit{E2F3} & \textit{TFDP1} & \textbf{\textit{BPTF}}\\
             \textit{WT1} & \cite{Rampal2016} R. Rampal, et al. (2016) & \textbf{\textit{SMAD4}} & \textit{SMAD5} & \textbf{\textit{HOXB6}}\\
             \textit{ZEB1} & \cite{Shousha2019} W. G. Shousha, et al. (2019) & \textit{MITF} & \textit{PAX3} & \textit{MLXIPL}\\
        \end{tabular}
    \end{ruledtabular}
\end{table*}

\bibliographystyle{unsrt} 
\bibliography{Triadic_bib}

\end{document}